\documentclass[a4paper,12pt,oneside]{article}
\usepackage[latin1]{inputenc}
\usepackage{amsmath}
\usepackage{amssymb}
\usepackage{amsfonts}
\usepackage{amssymb}
\usepackage{graphicx}
\usepackage{footnote}
\usepackage{float}
\usepackage{subcaption}
\DeclareGraphicsExtensions{.bmp,.png,.pdf,.jpg,.eps}
\usepackage{physics}
\usepackage{mathrsfs}

\usepackage[T1]{fontenc}
\usepackage{calligra}

\usepackage[colorlinks]{hyperref}
\hypersetup{
    colorlinks=true,
    linkcolor=blue,
    filecolor=magenta,  
    urlcolor=[rgb]{0.00,0.00,0.50},    
    citecolor=red,
    }

\usepackage{url}

\advance \textheight by \topskip
\usepackage{amsmath}
\usepackage[english]{babel}
\makeatletter
 \@addtoreset{equation}{section}
\makeatother
\usepackage[latin1]{inputenc}
\usepackage{amsmath}
\numberwithin{equation}{section}

\advance \textheight by \topskip
\usepackage{amsmath}
\usepackage[english]{babel}

\def\be{\begin{equation}}
\def\ee{\end{equation}}

\def\ba{\begin{eqnarray}}
\def\ea{\end{eqnarray}}

\def\Z{\mathbb Z}

\def\R{\mathbb R}

\usepackage{todonotes}

\begin{document}
\title{{\Large {\bf   \textcolor[rgb]{0.10,0.00,0.10}{Weak-strong duality of the non-commutative Landau problem
 induced by a two-vortex permutation,  and conformal bridge transformation}\\
[2pt]
 }}}

\author{{\bf  Andrey Alcala 
 and  Mikhail S. Plyushchay
 } 
 \\
[8pt]
{\small \textit{Departamento de F\'{\i}sica,
Universidad de Santiago de Chile,
Av. Victor Jara 3493, Santiago, Chile}}\\
[4pt]
 \sl{\small{E-mails:   
\textcolor{blue}{andrey.alcala@usach.cl},
\textcolor{blue}{mikhail.plyushchay@usach.cl}
}}
}
\date{}
\maketitle
\begin{abstract}
A correspondence is established between 
the dynamics of the two-vortex system  and 
the non-commutative Landau problem (NCLP) in its sub- (non-chiral), super- (chiral) 
and critical phases. 
As a result, a trivial permutation symmetry of the point
vortices induces a weak-strong coupling  duality in the NCLP.
We show  that   quantum two-vortex systems with 
non-zero total vorticity can be generated by applying 
conformal bridge transformation to a two-dimensional quantum 
free particle or to a quantum vortex-antivortex system 
of zero total vorticity.
The sub- and super-critical phases  of the quantum NCLP 
are generated in a similar way  from the  2D quantum free  particle  in a commutative 
or  non-commutative plane.  
The  composition of the inverse and direct transformations of the conformal 
bridge also makes it possible to link the non-chiral and chiral phases in 
each of these two systems.
\end{abstract}

\section{Introduction}     
Correspondences and connections  
between different systems, models and theories play an important role in physics.
An  example of this kind is the AdS/CFT correspondence or the Gauge/Gravity duality
\cite{ADS1}--\cite{ADS9}.
 On a simpler level an interesting example of the hidden correspondences between some integrable systems 
 is  provided by   the Newton-Hooke duality, based on conformal mapping \cite{Arnold}--\cite{InoJun},
  and its generalizations in the form of 
 the coupling constant metamorphosis \cite{HGDR,KMP}.
 In the same vein, the properties of many important physical systems
can be derived or explained by their hidden connection with the simplest 
system of a free particle. 
One of the striking examples of this  is the construction of 
(multi)soliton solutions of the classical KdV equation and equations 
of its hierarchy based on the Schr\"odinger problem for a one-dimensional 
free particle. 
The  construction, related to the inverse scattering method, 
 employs  the Darboux covariance of the corresponding  
 Lax pair formulation  for this integrable system \cite{MatSal,AraPlyu}.
 In a similar way, the B\"acklund transformations make 
 it possible to connect various integrable systems and generate 
 more complex solutions for them starting from simpler ones, 
 in particular, from trivial, identically equal to zero solutions  of the same  or 
related systems
 \cite{RogSch}.
 
 Some time ago, the construction of the  conformal bridge transformation 
 (CBT) made it possible to connect various harmonically confined 
 quantum mechanical systems 
 with a free (or asymptotically free) particle in spaces of various dimensions and geometric backgrounds, 
 including those  of a magnetic monopole and a cosmic string \cite{CBT1}--\cite{CBT4},
 see also refs. \cite{AchLiv1,AchLiv2}.
 In this way, explicit and hidden symmetries as well as  super-symmetries 
 of various systems can be derived  from a free particle. 
Also,   their eigenstates, coherent and squeezed states can be obtained 
 from certain states of the free particle quantum system.
 The CBT construction, based on a non-unitary 
 non-local similarity transformation, is analogous 
 to the Weierstrass transformation \cite{CBT1}. At the same time, it
 turns out to be closely related to the $\mathcal{PT}$-symmetry 
 \cite{Bender}
 through the underlying Dyson map \cite{Dyson}, 
 which transmutes the topological nature of the $\mathfrak{sl}(2,\R)$ 
 conformal symmetry generators \cite{CBT5,CBT6}.
 The non-local CBT generator has the nature of 
 the eighth-order  root of the identity  operator in the quantum phase 
 space and also turns out to be related to the unitary Bargmann-Segal 
 transformation, which connects the Hilbert space in the coordinate 
 representation with the Fock-Bargmann space of the holomorphic 
 representation of Heisenberg algebra \cite{CBT1}.
 
 The usual Landau problem, 
 also covered by the CBT \cite{CBT1,CBT7}, underlies the quantum Hall effect.
 The properties of the fractional quantum Hall effect can be explained 
 in terms of fractional statistics of the corresponding quasiparticle (anyon) 
 excitations \cite{ASW,NSSFS}.   Anyons  can be realized in the form 
 of point vortices through the mechanism 
 of statistics transmutation by using the Chern-Simons gauge theory \cite{LeiMyr,Wilc1}.
 Point vortices also appear in many non-linear field systems 
 and phenomena, including superconductivity, superfluidity, 
 and Bose-Einstein condensate physics \cite{Wilc2}--\cite{Penna}.
 Recently, the Landau problem has attracted attention in the
  study of the non-relativistic conformally invariant Schwartzian mechanical system 
  associated with the low energy limit of the Sachdev-Ye-Kitaev model 
  \cite{MalSta}--\cite{GanPol}.
 A  generalization of the usual Landau problem to 
 the case of noncommutative quantum mechanics 
 has been actively studied in the context of physics 
 associated with noncommutative geometry 
 \cite{LeinMyrh}--\cite{non-Lan7}.
\vskip0.1cm

The present work   is devoted to establishing a correspondence between
 the classical and quantum dynamics of 
 the simplest system of two point vortices and the
  non-commutative Landau problem  (NCLP) 
  in its sub- (non-chiral), super- (chiral), and critical phases.
This will be achieved by introducing an imaginary mirror particle into the
 latter  system. 
 The established correspondence will allow us to reveal a kind of the weak-strong coupling duality in
the NCLP induced by a trivial permutation symmetry of vortices.
By an appropriate generalization of the CBT construction
we  relate  the chiral and non-chiral quantum phases 
of both systems with a free particle in commutative and 
non-commutative plane. Using the same transformation, 
the quantum dynamics of the two-vortex systems with 
non-zero total vorticity will be related to a vortex-antivortex 
system of  zero total vorticity.

\vskip0.1cm

 The paper is organized as follows.
 In Section \ref{Section2} we briefly review the 
 symplectic structure and integrals of motion for the  general case
 of a system of $N$ point vortices. In Section \ref{Section3} we focus on  the 
case of superintegrable two-vortex systems. First we consider 
the two-vortex systems of non-zero total vorticity emphasizing  the important 
  difference  of their classical and quantum dynamics  in the cases of  the 
 vorticity strengths of the opposite and  equal signs.
 Then we consider  the 
 vortex-antivortex system of zero total vorticity with its 
emergent (1+1)D ``isospace" Lorentz  symmetry. 
 Section \ref{Section4}
 is devoted to the NCLP  in sub- (non-chiral),
 super- (chiral) and critical phases and to establishing   its correspondence with 
  the two-vortex systems
  by introducing an imaginary mirror particle into it.
   We show there that through the identified correspondence,  a
   trivial  permutational symmetry
 of  vortices induces a weak-strong coupling duality in the NCLP.
 In Section \ref{Section5} we study the conformal bridge transformation
 for the NCLP and two-vortex systems in chiral and non-chiral phases.
 To do this, we consider the integral of motion, 
 which is a certain linear combination of generators of 
 angular and time  translations
 in the NCLP.
 This integral, and its analog in the two-vortex system, 
 along with the integral of angular momentum, 
 will allow us to find two different canonical sets of variables 
 that have dual properties with respect to them  in the non-chiral and chiral phases. 
 Section \ref{Section6} is devoted to the summary, discussion and outlook.
In  Appendix, we discuss the basics of the simplest conformal bridge transformation, 
a generalization of which is used for the constructions in Section \ref{Section5}. 
There, we also provide an explanation for the difference between 
non-chiral and chiral phases for the systems under consideration 
in the light of the outer $\Z_2$ automorphism of the conformal 
$\mathfrak{sl}(2,\R)$ algebra, which has the nature of a
 $\mathcal{PT}$-inversion.

    
\section{N-vortex system and its symmetries}\label{Section2}

A system of $N\ge 2$  point vortices on a plane can be described by Lagrangian 
\cite{ArnoldKhesin,Newton}
\ba\label{NV-L}
&L=-\frac{1}{2}\sum_{a=1}^{N}\gamma_a\epsilon_{ij}x_a^i \dot{x}_{a}^j+ 
\frac{1}{2}\sum_{a<b}\gamma_a \gamma_b \log r_{ab}^2\,,&
\ea
where $\epsilon_{ij}$ is an antisymmetric tensor, $\epsilon_{12}=1$,
$\vec{x}_{a}$ 
are coordinates  of a vortex with index  $a$ and 
strength  $\gamma_a$, 
$\vec{x}_{a}\neq \vec{x}_{b}$  for $a\neq b$ $\Rightarrow r_{ab}^2=(\vec{x}_a- \vec{x}_b)^2>0$.
We do not distinguish between superscripts  and subscripts  of a 
two-dimensional space and imply summation over repeated indices.
 The vortex coordinates 
 and their strengths   $\gamma_a$ are assumed to be dimensionless. 
 
 It was already shown by Kirchhoff
 \cite{Kirchh}  that the Euler-Lagrange equations
  of motion 
  \ba\label{EOM-N}
&\dot{x}_{a}^{i}=-{\epsilon}^{ij} \sum_{b\neq a}\gamma_b{(x_{a}^{j}-x_{b}^{j})}r_{ab}^{-2}\,&
\ea
 admit a Hamiltonian description given by  
\ba\label{DPbr}
&\{x^i_{a},x^j_b\}=\gamma_{a}^{-1}\delta_{ab}\epsilon^{ij}\,,\qquad
H=-\frac{1}{2}\sum_{a<b}\gamma_a\gamma_b\log r_{ab}^2\,.&
 \ea
Symplectic structure with  Dirac-Poisson brackets (DPBs) of such a  form 
can also be obtained from Lagrangian of Landau problem 
by setting the mass parameter  to  zero  \cite{DJT}.

System  (\ref{NV-L})  is characterized by the Noetherian integrals of motion 
\ba\label{PiJ}
&P_i=\epsilon_{ij}\gamma_ax^j_a\,,\qquad M=-\frac{1}{2}\gamma_a\vec{x}_a^{\, 2}\,,&\label{PJint}
\ea
associated with
 translations,  $\{P_i,x^j_a\}=-\delta_{ij}$, and rotations, 
$\{M,x^a_i\}=\epsilon_{ij}x^j_a$.  They generate
a centrally extended Lie algebra   $\mathfrak{e}_\Gamma(2)$ of the two-dimensional 
Euclidean group,
\ba\label{e2G}
&\{P_i,P_j\}=\epsilon_{ij}\Gamma\,,\qquad
\{M,P_i\}=\epsilon_{ij}P_j\,,\qquad
\Gamma=\sum_a \gamma_a\,,&
\ea
with   total vorticity $\Gamma$ playing  a role of  a central charge,
and Casimir element  
\ba\label{Casi}
&\mathcal{C}_{\Gamma}=P_i^2+2\Gamma M=-\sum_{a< b}\gamma_a\gamma_b r_{ab}^2\,.&
\ea
These intergrals together with Hamiltonian generate the 
 $\mathfrak{e}_\Gamma(2)\oplus \mathfrak{u}(1)$ algebra, which reduces to 
 $\mathfrak{e}(2)\oplus \mathfrak{u}(1)$ when   $\Gamma=0$.
System  (\ref{DPbr})  is maximally superintegrable in the case of $N=2$ vortices,
completely integrable for $N=3$ (unlike 
the three-body problem of gravitating mass points \cite{Kirchh}), and 
 is not integrable, chaotic   for $N\geq 4$ \cite{ArnoldKhesin,Newton}.
In general case of $N$  vortices, equations of motion (\ref{EOM-N}) are also  invariant 
under  rescaling  $x^i_a\rightarrow e^\alpha x^i_a$, $t\rightarrow e^{2\alpha}t$. 
At the same time, Lagrangian is quasi-invariant,
$L\rightarrow L+\frac{d}{dt}(Ct)$, $C=\alpha \sum_{a<b}\gamma_a\gamma_b$,
but the action $S=\int Ldt$  is rescaled,   $S\rightarrow e^{2\alpha} S$, and
this symmetry of equations of motion  is not Noetherian, see also  \cite{Galazh}.


\section{Two-vortex system}\label{Section3}

In the case of the two-vortex superintegrable  system,   
Hamiltonian  (\ref{DPbr}) reduces to a function of  the Casimir element 
(\ref{Casi}) of  the $\mathfrak{e}_\Gamma(2)$ symmetry,
\be\label{H2vort}
H_\Gamma=-\frac{1}{2}\gamma_1\gamma_2\log r_{12}^2=
-\frac{1}{2}\kappa\log \left(-\kappa^{-1}\mathcal{C}_\Gamma\right)\,,\qquad
\kappa:=\gamma_1\gamma_2\,.
\ee
From now on, we will consider the two-vortex system.

\subsection{Nonzero total vorticity}

\begin{figure}[t!]
    \centering
    \begin{subfigure}[b]{0.19\textwidth}
           \centering
           \includegraphics[width=\textwidth]{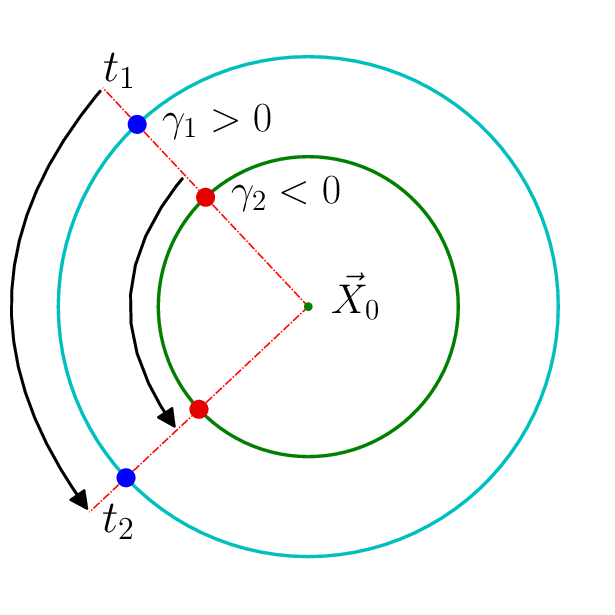}
            \caption{\small{$\kappa=\gamma_1\gamma_2<0$}}
            \label{fig:a}
    \end{subfigure}
    \qquad \quad\quad
    \begin{subfigure}[b]{0.19\textwidth}
            \centering
            \includegraphics[width=\textwidth]{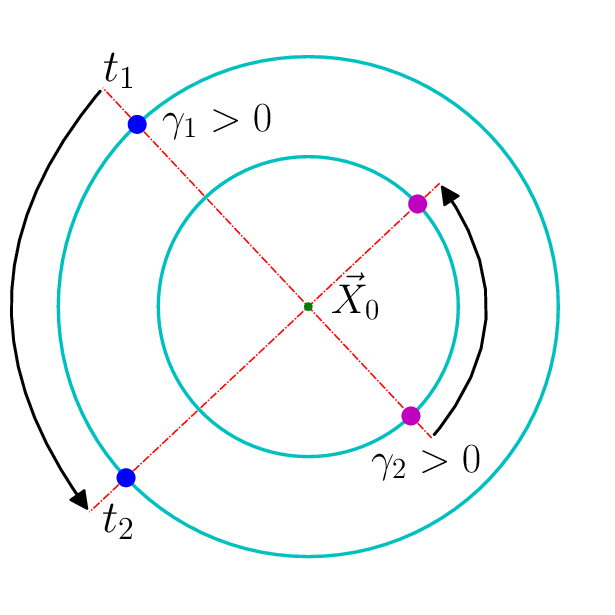}
            \caption{\small{$\kappa>0$, $\gamma_1\neq\gamma_2$}}
            \label{fig:b}
    \end{subfigure}
    \qquad \quad\quad
	\begin{subfigure}[b]{0.19\textwidth}
            \centering
            \includegraphics[width=\textwidth]{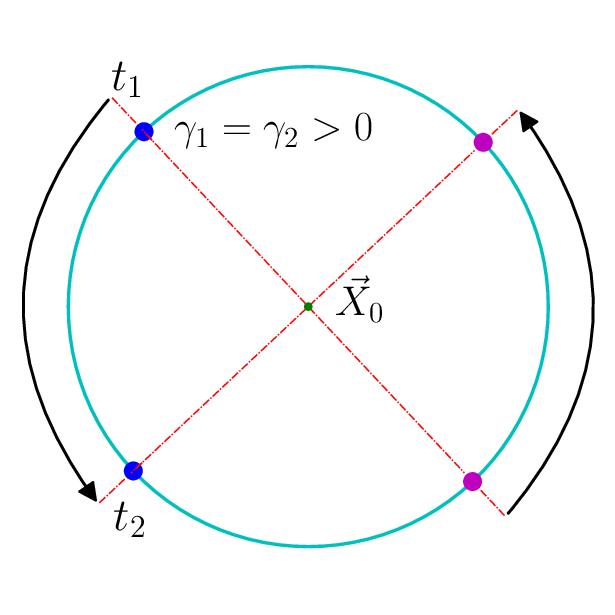}
            \caption{\small{$\gamma_1=\gamma_2$}}
            \label{fig:c}
    \end{subfigure}
    \caption{\small{Two-vortex dynamics with nonzero total vorticity   $\Gamma\neq 0$,  $t_2>t_1$.}}
    \label{Figure1}
\end{figure}

Let us first consider 
the case of nonzero total vorticity $\Gamma\neq 0$. It is convenient to describe 
the system 
by the vector integral 
$\vec{P}$  and 
the relative  coordinate vector $\vec{r}$, 
\ba\label{relcoor}
&P_i=\epsilon_{ij}\left(\gamma_1x^j_1+\gamma_2x^j_2\right)\,, \qquad
r^i_{12}=x^i_1-x^i_2:=r^i\,,\qquad r^ir^i>0\,,&\\
\label{DPBrP}
&\{r_i,r_j\}=\varrho^{-1}\epsilon_{ij}\,,\quad
\{P_i,P_j\}=\Gamma \epsilon_{ij}\,,\quad
\{r_i,P_j\}=0\,,\quad \varrho=
(\frac{1}{\gamma_1}+\frac{1}{\gamma_2})^{-1}
=
\kappa
({\Gamma})^{-1}
\,,\quad &
\ea
where   $\varrho$ 
 is a reduced vorticity.
 Equations of motion 
$\dot{P}_i=0\,,$ $\dot{r}_i=-\omega\epsilon_{ij}r_j\,,$
$\omega=\Gamma R^{-2}\,,$
$R^2=\vec{r}{\,}^2$,
lead to  a circular motion dynamics,
\ba
&x^i_a(t)=X^i_0+\Gamma^{-1}\epsilon_{ab}\gamma_b r^i(t)\,,\qquad X^i_0= -\Gamma^{-1}\epsilon^{ij}P_j \,,
&\label{orbit1}\\
&r_1(t)=R\cos\omega (t-t_0)\,,\quad
r_2(t)=R\sin\omega (t-t_0)\,,\qquad 
R=\exp\left(-\kappa^{-1}H_\Gamma\right)\,.\label{orbit2}&  
\ea  
For vorticities  of the opposite sign, $\kappa<0$,
the vortices occupy the closest positions on the circles with a common  ``center of vorticity"
$\vec{X}_0$,
while for $\kappa>0$ they are always  in the opposite  positions on their
circular orbits  of  radii  $R_1=\vert \gamma_2 \Gamma^{-1}\vert R$ 
and $R_2=\vert \gamma_1 \Gamma^{-1}\vert R$.
Identical vortices  with $\gamma_1=\gamma_2$ move on the same 
circle of radius $R/2$, see Fig. \ref{Figure1}.

Angular momentum  (\ref{PJint})
can be  expressed  in terms of the  integrals $P_i^2$ and $R^2$,
\ba\label{JP2R2}
&M_\Gamma=-\frac{1}{2\Gamma}\left(P_i^2+\kappa R^2\right)\,.&
\ea
It can take arbitrary values, $M_\Gamma\in \R$,  if $\kappa<0$,
but only nonzero values of the sign
$-\varepsilon_\Gamma=-\text{sgn}\,\Gamma$,  in the case 
of  $\kappa>0$. For $\kappa<0$, the angular momentum 
takes, in particular,  zero value, $M_\Gamma=0$,  
when 
$\vert\vec{X}_0\vert=\sqrt{R_1R_2}=\sqrt{\vert \varrho\vert}\, R$. 

Introduce now generators $\mathcal{A}^\pm$ and 
$\mathcal{B}^\pm$ of the two copies  of  classical Heisenberg algebra, 
$\{\mathcal{A}^-,\mathcal{A}^+\}=-i$, 
$\{\mathcal{B}^-,\mathcal{B}^+\}=-i$,
$\{\mathcal{A}^\pm,\mathcal{B}^\pm\}=0$, 
\ba\label{AABBdefin}
&\mathcal{A}^{-\varepsilon_{{\varrho}}}=\sqrt{\frac{|\varrho|}{2}}\,Z\,,\qquad
\mathcal{B}^{-\varepsilon_{\Gamma}}=\frac{1}{\sqrt{2|\Gamma|}}\,(P_1+iP_2)\,,\qquad
\varepsilon_{\varrho}=\text{sgn}\, \varrho\,,\quad
\varepsilon_\Gamma=\text{sgn}\, \Gamma\,,&
\ea
where $Z=e^{-i\omega t}z$,  $z=r_1+ir_2$,  and  $\bar{Z}$
are dynamical, explicitly depending on time, 
integrals of motion, $\frac{d}{dt}Z=\partial Z/\partial t +\{Z,H_\Gamma\}=0$.
Quadratic integrals constructed from them, 
\be\label{sl2sl2}
\mathcal{J}_0=\frac{1}{2}\mathcal{A}^+\mathcal{A}^-\,,\quad
\mathcal{J}_\pm=\frac{1}{2}\left(\mathcal{A}^\pm\right)^2\,,\qquad\quad
\mathcal{L}_0=\frac{1}{2}\mathcal{B}^+\mathcal{B}^-\,,\quad
\mathcal{L}_\pm=\frac{1}{2}\left(\mathcal{B}^\pm\right)^2\,,
\ee
generate  the algebra
$\mathfrak{sl}(2,\R)\oplus \mathfrak{sl}(2,\R)\cong AdS_2\oplus AdS_2\cong \mathfrak{so}(2,2)\cong 
AdS_3$,
\begin{eqnarray}\label{sl2(1)}
&\{\mathcal{J}_0,\mathcal{J}_\pm\}=\mp i\mathcal{J}_\pm\,,\qquad
\{\mathcal{J}_-,\mathcal{J}_+\}=-2i\mathcal{J}_0\,,&\\
&\{\mathcal{L}_0,\mathcal{L}_\pm\}=\mp i\mathcal{L}_\pm\,,\qquad
\{\mathcal{L}_-,\mathcal{L}_+\}=-2i\mathcal{L}_0\,,\qquad    
\{\mathcal{J}_{0,\pm},\mathcal{L}_{0,\pm}\}=0\,.\label{sl2(3)}&
\end{eqnarray}
Hamiltonian and angular momentum are represented in terms  
 of  the compact  generators $\mathcal{J}_0$ and $\mathcal{L}_0$ 
not depending explicitly on time, 
\ba\label{HJclasN=2}
&H_\Gamma=-\frac{1}{2}\kappa\log\left( 4|\varrho|^{-1} \mathcal{J}_0\right)\,,\qquad
M_\Gamma=-2\left(\varepsilon_\varrho\mathcal{J}_0+\varepsilon_{{}_\Gamma}\mathcal{L}_0\right)\,,&
\ea
and the  integral $\mathcal{J}_0$ is related, in turn,  to the Casimir element 
 (\ref{Casi}) of the  $\mathfrak{e}_\Gamma(2)$
symmetry, 
 $\mathcal{J}_0=\frac{1}{4}\vert \Gamma^{-1}\mathcal{C}_\Gamma\vert$.
Hamiltonian together with the dynamical integrals 
$\mathcal{J}_\pm$ generate a peculiar  nonlinear deformation
of the conformal  $\mathfrak{sl}(2,\R)$ algebra,
\ba\label{Hnonlinalg}
&\{\mathcal{J}_-,\mathcal{J}_+\}=\frac{1}{2}i\vert \varrho \vert R^2\,,\quad
\{H_\Gamma, \mathcal{J}_\pm\}=\pm 2i\,\varepsilon_{\kappa}
\vert \Gamma\vert R^{-2}\, 
\mathcal{J}_\pm \,,\quad
R^2= \exp(-2\kappa^{-1}H_\Gamma) \,.\quad  &
\ea

Dropping  the exponential time-dependent factors in 
$\mathcal{A}^\pm$, and changing  the  notation 
$\mathcal{A}^\pm\rightarrow a^\pm$
and  $\mathcal{B}^\pm\rightarrow b^\pm=\mathcal{B}^\pm$ in (\ref{AABBdefin}) and (\ref {sl2sl2}),
we do not modify  the algebraic relations (\ref{sl2(1)}), (\ref{sl2(3)}),
 and (\ref {Hnonlinalg}).  
Then, the  canonical quantization of the system entails the commutation relations 
$[\hat{a}^-,\hat{a}^+]=1$, $[\hat{b}^-,\hat{b}^+]=1$, 
$[\hat{a}^\pm,\hat{b}^\pm]=0$,   
 and the Fock space 
  is spanned by eigenstates $\ket{n_a,n_b}$
  of the number operators $\hat{N}_a=\hat{a}^{+}\hat{a}^{-} $,
  $\hat{N}_b=\hat{b}^{+}\hat{b}^{-} $ with eigenvalues $\,n_a,n_b=0,1,\ldots$.
The quantum analogs of (\ref{sl2sl2}) with 
antisymmetrized ordering in compact generators,
$\hat{\mathcal{J}}_0=\frac{1}{4}(\hat{a}^+\hat{a}^-
+\hat{a}^-\hat{a}^+)$,  
$\hat{\mathcal{L}}_0=\frac{1}{4}(\hat{b}^+\hat{b}^-
+\hat{b}^-\hat{b}^+)$,
yield  the quantum version of the $\mathfrak{sl}(2,\R)\oplus \mathfrak{sl}(2,\R)$ algebra
(\ref{sl2(1)}), (\ref{sl2(3)}).
In accordance with  (\ref{HJclasN=2}), 
quantum states $\ket{n_a,n_b}$   
are  eigenstates of the Hamiltonian and angular momentum operators,
\begin{eqnarray}\label{quantHG}
&\hat{H}_\Gamma\ket{n_a,n_b}=E_{n_a} \ket{n_a,n_b}\,,\qquad 
E_{n_a}=-\frac{1}{2}\kappa\log\left( 2 |\varrho|^{-1}\left(n_a+\frac{1}{2}\right) \right)\,,&\\
&\hat{M}_\Gamma\ket{n_a,n_b}=M_{n_a,n_b}\ket{n_a,n_b}\,,\qquad
M_{n_a,n_b}=-\varepsilon_{\varrho}\left(n_a+\frac{1}{2}\right)-\varepsilon_{\Gamma}\left(
n_b+\frac{1}{2}\right)\,.&\label{quantJG}
\end{eqnarray}
In the case of vorticities  of opposite sign, $\kappa=\gamma_1\gamma_2<0$,
the energy eigenvalues $E_{n_a}$ are bounded from below, while the angular momentum 
eigenvalues take integer values $M_{n_a,n_b}\in\Z$. 
For $\kappa>0$, the angular momentum 
takes nonzero integer values 
of the sign of $(-\Gamma)$, $ -\varepsilon_\Gamma M_{n_a,n_b}\in \Z_{>0}$,
whereas energy eigenvalues are bounded from above.   
Thus, the two-vortex system with  $\kappa<0$ is non-chiral,
while the dynamics of the sytem with $\kappa>0$ is chiral 
in the sense of  eigenvalues of the angular momentum operator.

\subsection{Zero total vorticity}

The case of zero total vorticity can be interpreted as the vortex-antivortex system.
It can be obtained from the $\kappa<0$, $\Gamma\neq 0$   case by taking the limit $\Gamma\rightarrow 0$.
For this, 
we denote $\gamma_1=\gamma$, $\gamma_2=-\gamma+\Gamma$, and  find
that the variables $\Pi_i$ and $\chi^i$ given by 
\ba\label{G0Pi}
&\lim_{\Gamma\rightarrow 0} P_i=\gamma\epsilon_{ij}r_j:=\Pi_i\,,\qquad 
\chi^i=\frac{1}{2}(x^i_1+x^i_2)\,, &
\ea 
form the canonical set of the variables, 
$\{\chi_i,\Pi_j\}=\delta_{ij}$,
$\{\chi_i,\chi_j\}=\{\Pi_i,\Pi_j\}=0$.
Hamiltonian (\ref{H2vort}) and angular momentum (\ref{JP2R2}) reduce  here  to
\ba\label{J0}
&\lim_{\Gamma\rightarrow 0}H_\Gamma=\frac{1}{2}\gamma^2\log (\gamma^{-2}\Pi_i^2):=H_0,\qquad
 \lim_{\Gamma\rightarrow 0}M_\Gamma =\epsilon_{ij}\chi_i\Pi_j:=M_0\,.&
\ea
The integrals $\Pi_i$,  $M_0$ and $H_0$  generate the $\mathfrak{e}(2)\oplus \mathfrak{u}(1)$ algebra
with  $\mathcal{C}_0=\Pi_i^2>0$  to be the Casimir of the 
$\mathfrak{e}(2)$ subalgebra. 

\begin{figure}[H] 
\begin{center}
\includegraphics[scale=0.35]{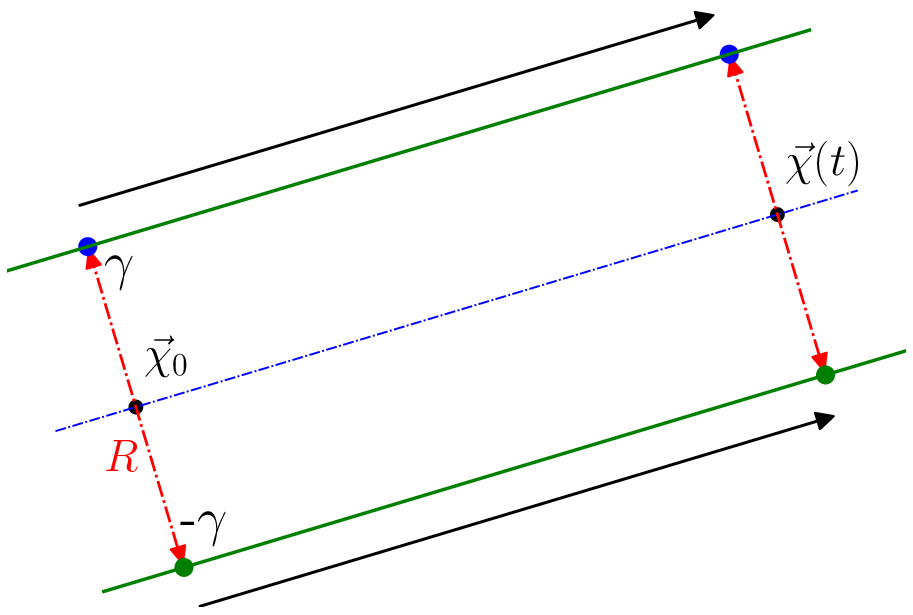}    
\caption{\small{The vortex-antivortex ($\Gamma=0$) dynamics.
}}
\label{Figure2}
\end{center}
\end{figure}

The equations of motion are $\dot{\Pi}_i=0$,  $\dot{\chi}_i=\gamma^2 \Pi_i/\vec{\Pi}^2$,
and the evolution of the system is 
\ba\label{xitpit}
&\chi^i(t)=\chi^i_0+\gamma^2\frac{\Pi^i}{\vec{\Pi}{}^2} t\,,\qquad
\Pi^i=const\,,&\\
\label{x(t)G0}
&x^i_1(t)=\chi^i_0-\frac{1}{2\gamma}\epsilon^{ij}\Pi_j +\gamma^2\frac{\Pi^i}{\vec{\Pi}{}^2} t\,,\qquad
x^i_2(t)=\chi^i_0+\frac{1}{2\gamma}\epsilon^{ij}\Pi_j +\gamma^2\frac{\Pi^i}{\vec{\Pi}{}^2} t\,,&
\ea
see  
Fig. \ref{Figure2}.
The distance between vortices is constant, $\vert \vec{r}\vert=\vert \vec{\Pi}/\gamma\vert=R$,
while the speed of the vortex and antivortex  is inverse to it,
$\vert \dot{\vec{\chi}}\vert =\vert \gamma\vert /R =\vert \gamma\vert \exp(-H_0/\gamma^2)$.
The rectilinear orbits of the vortex-antivortex system can be 
obtained by appropriately applying    the limit $\Gamma\rightarrow 0$ 
to the circular trajectories  of the system with $\kappa<0$ 
and nonzero total vorticity shown on panel (a) of Fig. \ref{Figure1}.  
For instance, by setting  $t_0=0$, $\gamma_1=\gamma$,
$\gamma_2=-\gamma +\Gamma$,  $P_1=\mu\Gamma R$, 
$P_2=\nu\Gamma R-\gamma$, $\mu, \nu\in \R$, 
and taking $\Gamma\rightarrow 0$,
we reduce 
Eqs. (\ref{orbit1}), (\ref{orbit2})   
to Eq. (\ref{x(t)G0}) 
 with
$\chi^1_0=(\frac{1}{2}-\nu)R$, $\chi^2_0=\mu R$, $\Pi_1=0$, $\Pi_2=-\gamma R$.

Dynamical scalar integrals 
\ba\label{Dint0}
&D=
{\chi}_0^i\Pi^i=
{\chi}^i {\Pi}^i- 
\gamma^2 t\,,\qquad
K=\frac{1}{2}{\vec{\chi}}^{{}^{\,\,2}}_0=\frac{1}{2}\vec{\chi}^{{}^{\,\,2}}-\left(\gamma^{-1}Dt+\frac{1}{2}
t^2\right)\exp\left(-2\gamma^{-2}H_0\right)\,\,\, \quad &  
\ea
together with the integral 
\ba\label{H0G0}
&\mathcal{H}_0:=\frac{1}{2}\vec{\Pi}^2=
\frac{1}{2}\gamma^2\exp(2H_0/\gamma^2)\,,&
\ea
generate the conformal $\mathfrak{sl}(2,\R)$ algebra,
\ba\label{DKH0}
&\{D,\mathcal{H}_0\}=2\mathcal{H}_0\,,\qquad
\{D,K\}=-2K\,,\qquad \{K,\mathcal{H}_0\}=D\,,&
\ea
whose Casimir is related to the angular momentum (\ref{J0}),
$\mathcal{C}=\mathcal{H}_0K-\frac{1}{2}D^2=\frac{1}{4}M_0^2$. 
With the change  $\mathcal{H}_0\rightarrow H_0$, the first and third relations in (\ref{DKH0})
are transformed into
\ba\label{DKH0def}
&\{D,H_0\}=\gamma^2\,,\qquad
\{K,H_0\}=D\exp\left(-2\gamma^{-2}H_0\right)\,.&
\ea
By dropping in (\ref{Dint0}) the terms that explicitly 
depend on time, i.e. changing there $\chi^i_0\rightarrow \chi^i$,    
the algebraic relations (\ref{DKH0}), (\ref{DKH0def}) are not modified.

Note that in the vortex-antivortex 
system ($\Gamma=0$), the first order in time derivative  term in Lagrangian  (\ref{NV-L})
with $N=2$ can be presented 
 as  $\frac{1}{2}\gamma\epsilon_{ij}\eta^{ab}x^i_a\dot{x}^j_b$,
where $\eta_{ab}=\text{diag}\,(-1,1)$
is the (1+1)D Minkowski metric in indexes $a,b=1,2$.
It is invariant under global (1+1)D Lorentz boost transformations of $x^i_a$ in 
``isospace" corresponding to index $a$, 
\ba\label{Lorentz}
&x'^i_1=\cosh\alpha\,x^i_1-\sinh\alpha\,x^i_2\,,\qquad
x'^i_2=\cosh\alpha\,x^i_2-\sinh{\alpha}\,x^i_1\,.&
\ea
They are generated by  the dynamical integral $D$, while 
the canonical variables $\Pi_i$ and $\chi_i$
undergo the rescaling transformations 
$\Pi_i'=e^{\alpha}\,\Pi_i$, $\chi_i'=e^{-\alpha}\,\chi_i$.
The linear in time $t$ term in integral $D$ 
 is associated 
with  quasi-invariance of the Lagrangian under   transformation (\ref{Lorentz}),
$ L'-L=\frac{d}{dt}(-\alpha\gamma^2 t)$. 
So,  the  vortex-antivortex system is characterized by the hidden
``isospace" (1+1)D Lorentz symmetry.
With the ``isospace"  Minkowski metric $\eta_{ab}$, 
the angular momentum integral (\ref{J0}) takes the form
$M_0=\frac{1}{2}\gamma\eta^{ab}\delta_{ij}x^i_ax^j_b$,
and the vortex-antivortex pair is described, in particular,
 by the 
$\mathfrak{so}(1,1)\oplus \mathfrak{so}(2)$ symmetry
generated by $D$ and $M_0$.
These two generators, along with $H_0$ and $K$  will play a key role 
in the CBT.

At the quantum level,  one can work   
in
representation with diagonal  operators $\hat{\chi}_j$. 
The plane wave functions 
$\psi_{{}_\Pi}(\xi)=\frac{1}{2\pi}\exp(i\chi_j\Pi_j)$, 
$\Pi_i^2>0$,    
are the normalized for delta function  eigenstates
 of $\hat{\Pi}_j=-i\,\partial/\partial {\chi}_j$, 
which are eigenstates of $\hat{H}_0$ of energy 
$E_{{}_\Pi}=\frac{1}{2}\gamma^2\log (\gamma^{-2}\Pi_i^2)\in \R $.
The angular momentum operator  takes integer eigenvalues $n\in \Z$.

\section{Non-commutative Landau problem}\label{Section4}

Let us compare now the two-vortex system 
with the non-commutative Landau problem (NCLP).
We take Lagrangian for the NCLP 
 in symmetric gauge in the form~\footnote{
 Various forms of the Lagrangian and symplectic structures
 are used 
 in the literature,  
 see, 
 e.~g.,  refs. \cite{non-Lan2+,HorDuv,non-Lan5}. 
 Our choice corresponds to
 refs.  \cite{HorDuv,non-Lan5} related to the exotic Galilean symmetry.}	
\begin{equation}\label{NCL-L}
\mathbb{L}= \mathbb{P}_i\dot{\mathbb{X}}_i + \frac{1}{2}\theta\epsilon_{ij}\mathbb{P}_i\dot{\mathbb{P}}_j+
\frac{1}{2}B\epsilon_{ij}\mathbb{X}_i\dot{\mathbb{X}}_j -\frac{1}{2m}\mathbb{P}_i^2\,,
\end{equation}
where $\mathbb{X}_i$ are the coordinates of the particle of  mass $m$ and charge  $e=1$ 
in magnetic field $B$,
 while $\theta$ is a non-commutativity parameter.
 The speed of light $c=1$.
 In the case $\beta:=B\theta\neq 1$, system (\ref{NCL-L}) 
 is described by  
 the Hamiltonian structure \cite{HorDuv,non-Lan5}  
 \vskip-0.5cm
\ba\label{DPBnL}
&\{\mathbb{X}_i,\mathbb{X}_j\}=\frac{\theta}{1-\beta}\epsilon_{ij}\,,\qquad
\{\mathbb{X}_i,\mathbb{P}_j\}=\frac{1}{1-\beta}\delta_{ij}\,,\qquad
\{\mathbb{P}_i,\mathbb{P}_j\}=\frac{B}{1-\beta}\epsilon_{ij}\,,&\\ 
\label{HmathP}
&\mathbb{H}=\frac{1}{2m}\mathbb{P}_i^2\,.&
\ea
We will see that 
the cases $\beta<1$ and $\beta>1$ correspond to non-chiral  
 and chiral  
 phases of the two-vortex system in spite of some differences in classical and quantum 
properties  of the two systems.
At the critical value $\beta=1$,  which separates the non-chiral and chiral phases,  
 DPBs (\ref{DPBnL}) blow up. 
 The Dirac-Bergmann analysis \cite{MarcClaudio}
 applied to the system (\ref{NCL-L}) shows that 
 at  $\beta=1$,  the corresponding  matrix of brackets of the constraints 
 degenerates, dimension of the reduced phase space decreases from four to two,
 and the system  is described by the coordinates $\mathbb{X}_i$ with DPBs
 $\{\mathbb{X}_i,\mathbb{X}_j\}=\theta\epsilon_{ij}$.
 Hamiltonian turns into zero,  $\mathbb{H}=0$ \cite{non-Lan2+,non-Lan3,non-Lan5}.
The spatial translation generators 
are $\mathcal{P}_i=\frac{1}{\theta}\epsilon_{ij}\mathbb{X}_j$,
while the angular momentum  $\mathbb{M}=-\frac{1}{2\theta}\mathbb{X}_i^2$ 
generates rotations of $\mathbb{X}_i$, and takes values on a half-line.
So the critical case $\beta=1$, which reveals a chiral nature,
 is essentially different from the cases 
with $\beta\neq 1$  because of 
a  trivial dynamics, 
$\dot{\mathbb{X}}_i=0$.
The angular momentum coincides here,
up to a sign, with Hamiltonian
of a 1D oscillator, which at the quantum level
takes half-integer values $M_n=-\varepsilon_\theta(n+\frac{1}{2})$, $\varepsilon_\theta=\text{sgn}\, \theta$,
 $n=0,1,\ldots$. 

The critical case $\beta=1$ of the NCLP  
corresponds to the two-vortex 
system with one of  the parameters  $\gamma_a$ set equal  to  zero, i.e.,  to a single-vortex system.
Indeed, putting there $\gamma_2=0$,  
we obtain  the second-class,  
$\phi_1^i=p^i_1-\frac{1}{2}\gamma_1 \epsilon_{ij}x^j_1\approx 0$, and  the  first-class, 
$\phi_2^i=p^i_2\approx 0$, constraints 
 with respect to the symplectic structure $\sigma=dp^i_a\wedge dx^i_a$ in the initial phase space.
 Constraints $\phi_2^i \approx 0$  mean that the variables  $x^i_2$ are of a pure 
gauge nature, and one
 can forget about the degrees of freedom corresponding to  index $a=2$.
Reduction to the subspace of the  constraints 
$\phi_1^i\approx 0$ results   in the two-dimensional symplectic manifold described
by coordinates  $x^i_1$ with the DPBs
$\{x_1^i,x_1^j\}=\gamma_1^{-1}\epsilon^{ij}$ and 
zero Hamiltonian. 
Identifying  $\theta=\gamma_1^{-1}$ and $\mathbb{X}^i=x^i_1$,
 we conclude  that the 
critical case of the NCLP  
 is 
equivalent to the reduced, single-vortex system
with a trivial  dynamics  
and a chiral nature of the angular momentum. 
In what follows we assume that $\beta\neq 1$.

\subsection{Non-critical phases} \label{Non-crit}  
Equations of motion generated by Hamiltonian 
(\ref{HmathP})  through DPBs (\ref{DPBnL}) are
\ba\label{eqXP}
&\dot{\mathbb{X}}_i=\frac{1}{m(1-\beta)}\mathbb{P}_i\,,\qquad
\dot{\mathbb{P}}_i=\frac{B}{m(1-\beta)}\epsilon_{ij}\mathbb{P}_j\,.&
\ea
The vector integral $\mathcal{P}_i=\mathbb{P}_i-B\epsilon_{ij}\mathbb{X}_j$ with DPBs
\ba\label{PPB}
&\{\mathcal{P}_i,\mathcal{P}_j\}=-B\epsilon_{ij}\,,\qquad
\{\mathbb{X}_i,\mathcal{P}_j\}=\delta_{ij}\,,\qquad
\{\mathbb{P}_i,\mathcal{P}_j\}=0\,,&\label{PPB2}
\ea
generates 
translations of $\mathbb{X}_i$.
Solutions to Eqs. (\ref{eqXP}) are 
\ba
&\mathbb{X}^i(t)=\mathbb{X}_0^i+ \mathcal{R}\,n^i(t)\,,\qquad
\mathbb{X}_0^i=\frac{1}{B}\epsilon^{ij}\mathcal{P}_j\,,\qquad  \mathcal{R}=\frac{\sqrt{2m\mathbb{H}}}{B}\,,&\\
&n^i(t)=(\cos(\omega_*(t-t_0)),\,\sin(\omega_*(t-t_0)))\,,\qquad
\omega_*=\frac{B}{m(1-\beta)}  \,.&
\ea
Like in the two-vortex system, the 
angular momentum integral 
\be\label{Meijqipj}
\mathbb{M}=\epsilon_{ij}\mathbb{X}_i \mathbb{P}_j+\frac{1}{2}
\left(\theta \mathbb{P}_i^2+B\mathbb{X}_i^2\right)=\frac{1}{2B}\left(\mathcal{P}_i^2
- (1-\beta)\mathbb{P}_i^2\right)
\ee
together with $\mathcal{P}_i$ generate 
the 
centrally extended Lie algebra   of the two-dimensional 
Euclidean group of the form (\ref{e2G})
with $-B$ playing the role of the central charge.
The Hamiltonian is presented  in terms of its Casimir element, cf.  Eq. (\ref{Casi}), 
\ba
&\mathbb{H}=\omega_*\left(\frac{1}{2B}\mathcal{P}_i^2-\mathbb{M}\right)\,.&
\ea

Let us define now the coordinate vector of an
``imaginary mirror particle",
\be\label{YdefX}
\mathbb{Y}_i=\mathbb{X}_i+\theta\epsilon_{ij}\mathbb{P}_j=
(1-\beta)\mathbb{X}_i+\theta\epsilon_{ij}\mathcal{P}_j\,,
\ee
which is characterized  by  three important properties\,:
it has zero DPBs with $\mathbb{X}_i$, 
 it  is translated by  $\mathcal{P}_i$,
 and the DPBs betwen its components do not depend on $B$,
\be\label{YYDPBs}
\{\mathbb{Y}_i,\mathbb{Y}_j\}=-\theta\epsilon_{ij}\,,\qquad
\{\mathbb{Y}_i,\mathbb{X}_j\}=0\,,\qquad
\{\mathbb{Y}_i,\mathcal{P}_j\}=\delta_{ij}\,.
\ee
In terms of 
$\mathbb{X}_i$ and $\mathbb{Y}_i$
the Hamiltonian and angular momentum are expressed as 
\be\label{HMXY}
\mathbb{H}=\frac{1}{2m\theta^2}(\mathbb{X}_i-\mathbb{Y}_i)^2\,,\qquad
\mathbb{M}=\frac{1}{2\theta}\left(\mathbb{Y}_i^2-(1-\beta)\mathbb{X}_i^2\right)\,,
\ee
and we also have
\ba\label{PPXY}
&\mathcal{P}_i=\frac{1}{\theta}\epsilon_{ij}\left((1-\beta)\mathbb{X}_j-\mathbb{Y}_j\right)\,,
\qquad
\mathbb{P}_i=\frac{1}{\theta}\epsilon_{ij}\left(\mathbb{X}_j-\mathbb{Y}_j\right)\,,&\\
\label{XYrelPP}
&\mathbb{X}_i=\frac{1}{B}\epsilon_{ij}\left(\mathcal{P}_j-\mathbb{P}_j\right)\,,
\qquad
\mathbb{Y}_i=\frac{1}{B}\epsilon_{ij}\left(\mathcal{P}_j-(1-\beta)\mathbb{P}_j\right)\,.&
\ea
Up to a total  time derivative, 
 Lagrangian 
(\ref{NCL-L})  takes the form similar to (\ref{NV-L}) at $N=2$, 
\be\label{LYLag}
\mathbb{L}=\frac{1}{2\theta}\epsilon_{ij}\left((\beta-1)\mathbb{X}_i\dot{\mathbb{X}}_j
+\mathbb{Y}_i\dot{\mathbb{Y}}_j\right)-\frac{1}{2m\theta^2}(\mathbb{X}_i-\mathbb{Y}_i)^2\,.
\ee

The four-dimensional phase space  
 can be described, up to a canonical transformation, 
   in terms of any pair of the  vector variables  $\mathbb{X}_i,$ $\mathbb{P}_i$, $\mathbb{Y}_i$
   and $\mathcal{P}_i$.
    The Hamiltonian and angular momentum take the simplest form  
    in terms of the pairs 
     $(\mathbb{P}_i, \mathcal{P}_i)$ and 
      $(\mathbb{X}_i, \mathbb{Y}_i)$.

Comparing the NCLP  with
the two-vortex system, 
one can establish, up to a permutation $1\leftrightarrow 2$ of the values of
index $a$,   the following correspondence between them\,:
\ba
&\mathbb{X}^i\sim x^i_1\,,\quad
\mathbb{Y}^i\sim x^i_2\,,\qquad
\mathbb{P}_i\sim -\gamma_2\epsilon_{ij}r^j\,,\qquad
\mathcal{P}_i\sim P_i\,,\qquad
\mathbb{M}\sim M_{\Gamma}\,,
&\label{corr1}\\
&(1-\beta)\theta^{-1}\sim\gamma_1\,,\qquad
{\theta}^{-1}\sim-\gamma_2\,,\qquad
B\sim-\Gamma\,,
\qquad
(1-\beta)\sim -\gamma_1\gamma_2^{-1}\,.&\label{corr2}
\ea
The difference of the two systems is reflected 
in a linear dependence of $\mathbb{H}$ on $\mathbb{P}_i^2$ or
$(\mathbb{X}_i-\mathbb{Y}_i)^2$, but logarithmic dependence of
$H_\Gamma$ on $(x^i_1-x^i_2)^2=R^2$.
Because  of this difference the  signed   frequency of rotational motion 
 in  the NCLP   
is given in terms of the  parameters 
$\theta$, $B$ and $m$, 
while 
in the two-vortex system 
the signed frequency is energy-dependent.
According to  (\ref{corr2}), 
the subcritical phase $\beta<1$ corresponds to the case of vorticities of
the opposite sign,  $\gamma_1\gamma_2<0$. 
In particular, the vortex-antivortex system, $\Gamma=0$,  is similar  to
 a free particle in non-commutative plane, $B=0$.
 The supercritical  phase with
$\beta>1$ corresponds to the chiral case of vorticities of the same sign,
$\gamma_1\gamma_2>0$, 
see Eqs (\ref{Meijqipj}), 
 (\ref{HMXY}).

The case of the ordinary  Landau problem, $\theta=0$, $B\neq 0$,
also can be  covered  by correspondence relations (\ref{corr1}), (\ref{corr2}).
For this,  note that in the limit  $\theta 	\rightarrow 0$, we get
$\mathbb{Y}_i=\mathbb{X}_i$,
$\mathbb{M}=\epsilon_{ij}\mathbb{X}_i\mathbb{P}_j+\frac{1}{2} B\mathbb{X}_i^2$,
$\{\mathbb{X}_i,\mathbb{X}_j\}=0$.
In the two-vortex system we 
 change the variables,  $(x_1^i, x^i_2) \rightarrow (x^i,\pi^i)$,
$x^i=x_1^i$, $x_2^i=x^i-\frac{1}{\gamma_2}\epsilon_{ij}\pi_j$,
present  
$\gamma_1=-\gamma_2+\Gamma$,
and redefine Hamiltonian by shifting  and rescaling it,  
$H_\Gamma \rightarrow h_\Gamma=
\frac{\Gamma^2}{\gamma_2^2}(H_\Gamma-
\frac{1}{2}\gamma_1\gamma_2\log\gamma_2^2 -\frac{1}{2}\gamma_2^2\log \Gamma^2)$. 
According to  (\ref{corr1}), (\ref{corr2}), 
 the ordinary Landau problem 
corresponds to the limit $\gamma_2 \rightarrow\infty$  in the two-vortex system.
This yields 
$x_2^i=x^i_1=x^i\sim \mathbb{X}^i$, 
$\pi_i\sim \mathbb{P}_i$, $M_\Gamma=\epsilon_{ij}x_i\pi_j-\frac{1}{2}\Gamma x_i^2\sim \mathbb{M}$,
$P_i=\pi_i+\Gamma\epsilon_{ij}x_j\sim \mathcal{P}_i$, $\{x_i,x_j\}=0$, $\{x_i,\pi_j\}=\delta_{ij}$,
$\{\pi_i,\pi_j\}=-\Gamma\epsilon_{ij}$. 
The dynamics 
is given by  the  ``regularized" Hamiltonian
$h_\Gamma=\frac{1}{2}\Gamma^2\log (\pi_i^2/\Gamma^2)$.  
 In correspondence with 
Eqs. (\ref{orbit1}), (\ref{orbit2}), $x^i(t)$ describes 
rotational motion with angular velocity 
$\omega=\Gamma/R^2$ along the circle of radius $R=
(\vec{\pi}^2/\Gamma^2)^{1/2}$  
centered at $X^i_0=-\frac{1}{\Gamma}\epsilon_{ij}P^j$.

Taking into account correspondence 
 (\ref{corr1}),  (\ref{corr2}), we find the analogs of the variables $a^\pm$ and $b^\pm$ of the two-vortex system,
\begin{eqnarray}\label{mathab1}
&\mathfrak{a}^{\varepsilon_\mathfrak{a}}=i\,\text{sgn}\,\theta\cdot 
\sqrt{\frac{\vert 1-\beta\vert}{2\vert B\vert}}
(\mathbb{P}_1+i\mathbb{P}_2)\,,\qquad
\mathfrak{b}^{\varepsilon_\mathfrak{b}}=\frac{1}{\sqrt{2\vert B\vert}}(\mathcal{P}_1+i\mathcal{P}_2)\,,&
\ea
where $\varepsilon_\mathfrak{a}=\text{sgn}\,(B(\beta-1))$, 
$\varepsilon_\mathfrak{b}=\text{sgn}\,B$.
The  analogs of quadratic quantities 
(\ref{sl2sl2}) given in terms of the variables $\mathfrak{a}^\pm$ and $\mathfrak{b}^\pm$
yield us the $\mathfrak{sl}(2,\R)\oplus \mathfrak{sl}(2,\R)\cong
AdS_3$ algebra of the 
non-commutative Landau problem,
and here
\ba
&\mathbb{H}=\vert \omega_*\vert\,  \mathfrak{a}^+\mathfrak{a}^-\,,\qquad
\mathbb{M}=\varepsilon_\mathfrak{a}\, \mathfrak{a}^+\mathfrak{a}^- +
\varepsilon_\mathfrak{b\, }\mathfrak{b}^+\mathfrak{b}^-\,.&
\ea

By analogy with 
the two-vortex system,
the quantum states $\ket{n_\mathfrak{a},n_\mathfrak{b}}$, $n_\mathfrak{a},n_\mathfrak{b}=0,1,\ldots$, are eigenstates
of the operators $\hat{\mathfrak{a}}^+\hat{\mathfrak{a}}^-$
and $\hat{\mathfrak{b}}^+\hat{\mathfrak{b}}^-$
with eigenvalues $n_\mathfrak{a}$ and $n_\mathfrak{b}$. They 
are  eigenstates of the energy
$E_{n_\mathfrak{a}}= \vert \omega_* \vert(n_\mathfrak{a}+\frac{1}{2})$ 
and  angular momentum 
$M_{n_\mathfrak{a},n_\mathfrak{b}}=\varepsilon_\mathfrak{a}(n_\mathfrak{a}+\frac{1}{2})
+\varepsilon_\mathfrak{\mathfrak{b}}(n_\mathfrak{a}+\frac{1}{2})$.
Thus, in the supercriticial (chiral) phase $\beta>1$,
the angular momentum takes nonzero integer values of the sign of 
magnetic field $B$, $\text{sgn}\, B\cdot  M_{n_\mathfrak{a},n_\mathfrak{b}}\in  \, \Z_{>0}$,
while in the subcritical phase $\beta<1$ it 
takes integer values of both signs, including zero value,
$M_{n_\mathfrak{a},n_\mathfrak{b}}\in \, \Z$,  in correspondence 
with the analogous picture in the two-vortex system.

The case of a free particle in non-commutative plane, $B=0$, $\theta\neq 0$, 
is similar to the vortex-antivortex system 
with zero total vorticity $\Gamma=0$.
For $B=0$, we have the vector $\mathbb{P}_i=\mathcal{P}_i\sim\Pi_i$, 
and the vector coordinate 
\ba\label{X+Ychi}
\mathcal{X}_i=\frac{1}{2}(\mathbb{X}_i+
\mathbb{Y}_i )=\mathbb{X}_i+\frac{1}{2}\theta \epsilon_{ij}\mathcal{P}_j\, \sim\,
\chi_i,&
\ea
see Eq. (\ref{G0Pi}), which 
form a canonical set of variables,
\ba\label{HJ0m}
&\{ \mathcal{X}_i,\mathcal{X}_j  \}=
\{ \mathcal{P}_i,\mathcal{P}_j \}=0\,,\qquad
\{ \mathcal{X}_i,\mathcal{P}_j \}=\delta_{ij}\,.&
\ea
With the Hamiltonian and angular momentum
\ba\label{B=0HJ}
&
\mathbb{H}_0=\frac{1}{2m}\mathcal{P}_i^2\,,\qquad
\mathbb{M}_0=\epsilon_{ij}\mathcal{X}_i\mathcal{P}_j\,,\qquad (B=0)\,,&
\ea
the system looks like a free scalar particle in
two-dimensional Euclidean space.
However, it is necessary to  bare in mind that the coordinates $\mathbb{X}_i$ of a free 
particle in non-commutative plane, $\{ \mathbb{X}_i,\mathbb{X}_j\}=\theta\epsilon_{ij}$,
are given  by 
$\mathbb{X}_i=\mathcal{X}_i-\frac{1}{2}\theta \epsilon_{ij}\mathcal{P}_j$. 
They form a 2D vector with respect to the angular momentum 
$\mathbb{M}_0=\epsilon_{ij}\mathbb{X}_i\mathcal{P}_j+
\frac{1}{2}\theta\mathcal{P}_i^2$,
and covariantly    transform  under the exotic Galilean boosts,
\be
\{\mathbb{X}_i,\mathbb{G}_j\}=-\delta_{ij} t\,,\qquad
\mathbb{G}_i=m\mathbb{X}_i-t\mathcal{P}_i
+m\theta\epsilon_{ij}\mathcal{P}_j\,,\qquad
\{\mathbb{G}_i,\mathbb{G}_j\}=m^2\theta\epsilon_{ij}\,,
\ee
where $\mathbb{G}_i$ is a dynamical integral. 
Unlike $\mathbb{X}_i$, 
the 2D vector $\mathcal{X}_i$
transforms non-covariantly under the exotic Galilean boosts,
$\{\mathcal{X}_i, \mathbb{G}_j\}=-\delta_{ij}t-\frac{1}{2}m\theta\epsilon_{ij}$,
and plays a role 
analogous to the Newton-Wigner coordinates 
for a Dirac particle~\footnote{
For discussion of the exotic Galilean 
symmetry of a particle in non-commutative plane 
and the aspects related to different choices of the coordinate 
variables see refs. \cite{non-Lan1,non-Lan4,JackNair,AnyNCP}.}. The 
coordinates  $\mathbb{Y}_i$   of imaginary particle
 also transform  non-covariantly under 
 the exotic Galilean boosts,
 $\{\mathbb{Y}_i,\mathbb{G}_j\}=-\delta_{ij}t-m\theta\epsilon_{ij}$.

As $\mathcal{X}^i(t)=\mathcal{X}^i_0+\frac{1}{m}\mathcal{P}^it$,
one can construct the quadratic dynamical integrals 
\ba
&\mathbb{D}=\mathcal{X}^i_0\mathcal{P}^i=
\mathcal{X}^i\mathcal{P}^i-2\mathbb{H}_0t\,,\qquad
\mathbb{K}=\frac{m}{2}\mathcal{X}^i_0\mathcal{X}^i_0=
\frac{m}{2}\mathcal{X}^i\mathcal{X}^i-\mathbb{D}t-
\mathbb{H}_0t^2\,,&
\ea
which together with the Hamiltonian $\mathbb{H}_0$
generate the conformal $\mathfrak{sl}(2,\R)$ algebra
of the form (\ref{DKH0}).
The phase space function
$\tilde{\mathbb{D}}=\mathcal{X}_i\mathcal{P}_i=
-\frac{1}{\theta}\epsilon_{ij}\mathbb{X}_i\mathbb{Y}_j$
together with $\tilde{\mathbb{K}}=\frac{m}{2}\mathcal{X}_i^2$ and 
$\mathbb{H}_0$ generate the same  $\mathfrak{sl}(2,\R)$ algebra.
Combining the 
coordinates $\mathbb{X}^i$ and $\mathbb{Y}^i$ into the 
``isospace"  vector
$\mathbb{X}_{\mathfrak{c}}^i=(\mathbb{X}^i,\mathbb{Y}^i)$,
we find that in index $\mathfrak{c}=1,2$ it behaves like a (1+1)-dimensional Lorentz vector
with respect to the global transformations generated by
$\tilde{\mathbb{D}}$, cf. (\ref{Lorentz}),
\ba
&\mathbb{X}'^i=\cosh{\alpha}\,\mathbb{X}^i-\sinh{\alpha}\,\mathbb{Y}^i\,,\qquad
\mathbb{Y}'^i=\cosh{\alpha}\,\mathbb{Y}^i-\sinh{\alpha}\,\mathbb{X}^i\,.&
\ea

 At the quantum level, the non-covariant with respect to the exotic Galilean boosts vector  $\mathcal{X}_i$ allows us to
work in representation diagonal in
$\hat{\mathcal{X}}_i$,
in which $\hat{\mathcal{P}}_j=-i\partial/\partial \mathcal{X}_j$,
$\hat{\mathbb{X}}_j=\mathcal{X}_j+i\frac{1}{2}\theta \, \epsilon_{jk}
\partial/\partial \mathcal{X}_k$, and 
 $\hat{\mathbb{Y}}_j=\mathcal{X}_j-i\frac{1}{2}\theta \,\epsilon_{jk}
\partial/\partial \mathcal{X}_k$.
Similarly to the vortex-antivortex system,  the wave function 
$\psi_{\mathcal{P}}(\mathcal{X})=\frac{1}{2\pi}\exp(i \mathcal{X}_j\mathcal{P}_j)$
is the eigenfunction of the Hamiltonian 
$\hat{\mathbb{H}}_0$ of  eigenvalue $E_{\mathcal{P}}=\frac{1}{2m}\mathcal{P}_i^2$,
and  $\hat{\mathbb{M}}_0$ takes integer eigenvalues $n\in\Z$.


\subsection{Vortex permutation 
and weak-strong duality of the NCLP}

In the two-vortex system, the 
transformation
$\mathcal{I}:\,\,\, x^i_1\leftrightarrow x^i_2,\,, 
\gamma_1\leftrightarrow \gamma_2$,
is the obvious symmetry of the nature of inversion, $\mathcal{I}^2=id$.
It acts as a rotation by $\pi$  on the relative coordinate $r^i=x^i_1-x^i_2$,
$\mathcal{I}:\,r^i\rightarrow -r^i$,
but
does not change the energy, the momentum vector  and the angular momentum integrals.
 In the NCLP,   the 
correspondence (\ref{corr1}), (\ref{corr2}) 
leads to the analog of the permutation of vortices,  
\ba\label{inversionLan}
&\mathcal{I}:\,\,\,  \mathbb{X}_i\leftrightarrow \mathbb{Y}_i\,,\quad   B\rightarrow B\,,\quad
\theta\rightarrow \frac{\theta}{\beta -1}\,,\quad  (1-\beta)\rightarrow (1-\beta)^{-1}\,,\quad
\quad \mathcal{I}^2=id\,.&
\ea
Under transformation (\ref{inversionLan}) the first  brackets in (\ref{DPBnL}) and
(\ref {YYDPBs}) transform one into another, and the corresponding symplectic two-form
$\sigma=-\frac{1}{2\theta}\epsilon_{ij}((1-\beta)d\mathbb{X}_i\wedge d\mathbb{X}_j
+d\mathbb{Y}_i\wedge d\mathbb{Y}_j)$ is invariant.
The integral  $\mathcal{P}_i$ and the angular momentum $\mathbb{M}$ 
are invariant under transformation (\ref{inversionLan}),
but  $ \mathbb{P}_i$ and $ \mathbb{H}$ are rescaled,
\ba
&\mathcal{I}:\,\,\, \mathbb{P}_i\rightarrow (1-\beta)\,\mathbb{P}_i\,,\qquad
\mathbb{H}\rightarrow (1-\beta)^{2}\, \mathbb{H}\,.&
\ea
The Hamiltonian $\mathbb{H}$ 
is invariant under 
transformation (\ref{inversionLan}) only when $\beta=0$ or $\beta=2$. 
The first case corresponds either to  a free particle in non-commutative plane ($B=0$, $\theta\neq 0$),
which 
is  analogous to the vortex-antivortex system with $\Gamma=0$,
or to the ordinary Landau problem ($\theta=0$, $B\neq 0$).
The second case with $\theta=2/B$ is similar to the  system of identical vortices  with
$\gamma_1=\gamma_2$. 
For  $\beta\neq 0, 2$, 
the transformation relates the modes  of a weak and strong coupling
  (in the sense of energy levels spacing 
$\Delta E=\vert \omega_*\vert$)
of the NCLP with respect to the critical value $\beta=1$ within the same 
sub- or super- critical phase,  $\mathcal{I}:\,\,\,[(\beta-1) \rightarrow 0^\pm]\,\, \longleftrightarrow\,\,
[(\beta-1)\rightarrow \pm \infty]$.  
As $\mathcal{I}^2=id$, 
transformation (\ref{inversionLan}) can be treated as a kind of  a weak-strong coupling duality.


\section{Conformal bridge}\label{Section5}

In this section we show how the quantum NCLP
can be generated from a free particle system 
in a commutative or non-commutative plane
by means of the conformal bridge transformation \cite{CBT1,CBT5,CBT6}.
In the sub- ($\beta<1$) and super- ($\beta>1$) 
critical phases the symplectic structure of the 
system is distinct. Though the components of 
non-commutative coordinate $\mathbb{X}_i$ 
and  momentum $\mathbb{P}_i$ form the 2D 
vectors with respect to the angular momentum $\mathbb{M}$, 
all their brackets among themselves  
have different signs in these two 
phases while the signs of the magnetic field $B$ and the 
non-commutativity parameter $\theta$ are maintained fixed.
Moreover, their brackets blow up at $\beta=1$, 
and the system decreases  the dimension 
of the  phase space from four to two. 
Coherently with these changes,  
in the
 supercritical phase the system acquires chirality: 
 the angular momentum  integral $\mathbb{M}$  takes 
 values of  only one sign. 
 As a consequence, generation of  the two phases of the 
 NCLP   
  from a free particle  is carried out in 
  distinct  ways. The reason is  that in supercritical case 
the chiral nature of the angular momentum is incompatible 
with vector properties of Darboux coordinates. 
 The correspondence between the NCLP    and 
the two-vortex system  allows then to generate the latter system from 
the free particle in a plane separately in the cases of vorticities 
$\gamma_a$ of the opposite  or  equal signs. 
 
 \vskip0.1cm
 
 In connection  with appearance of chirality, 
we consider one more integral 
$\breve{\mathbb{M}}:=\frac{1}{2B}(\mathcal{P}_i^2+(1-\beta)\mathbb{P}_i^2)$,  
which  
  is 
 a linear combination of  
  the angular momentum 
 and the Hamiltonian,
 \ba\label{breveM}
& \breve{\mathbb{M}}=\mathbb{M}+\frac{2m(1-\beta)}{B}\mathbb{H}\,.&
 \ea 
  We also introduce 
 the two-component object $\check{\mathbb{P}}_i$ obtained from  $\mathbb{P}_i$ 
 by a spatial reflection, 
\ba\label{widetildePdef}
&\check{\mathbb{P}}_i:=(-\mathbb{P}_1,\mathbb{P}_2)\,,\qquad
\check{\mathbb{P}}_i{}^2=\mathbb{P}_i^2.&
\ea
 One finds then that the  integral $\mathcal{P}_i$ is transformed as a vector  by both integrals 
$\mathbb{M}$ and $\breve{\mathbb{M}}$, while 
$\mathbb{P}_i$ and $\check{\mathbb{P}}_i$ are transformed as 
2D vectors only by $\mathbb{M}$ and $\breve{\mathbb{M}}$, respectively:
 \ba\label{MVbra}
 &\{\mathbb{M},\mathcal{V}_i\}=s\epsilon_{ij}\mathcal{V}_j\,, \quad
 \qquad s=(+,+,-)\,\, \text{for}\,\, \mathcal{V}_i=(\mathcal{P}_i,\mathbb{P}_i,
 \check{\mathbb{P}}_i)\,,&\\ 
 \label{tilMVbra}
 &\{\breve{\mathbb{M}},\mathcal{V}_i\}=\breve{s}\epsilon_{ij}\mathcal{V}_j\,, \quad
 \qquad \breve{s}=(+,-,+)\,\, \text{for}\,\, \mathcal{V}_i=(\mathcal{P}_i,\mathbb{P}_i,
\check{\mathbb{P}}_i)\,.&
 \ea

\subsection{Canonical variables for two phases of the NCLP}

Consider now  the subcritical phase $\beta<1$ of the 
NCLP. 
One can define 
canonical (Darboux) variables $(\mathfrak{q}_i,\mathfrak{p}_i)$, 
$
\{\mathfrak{q}_i,\mathfrak{q}_j\}=\{\mathfrak{p}_i,\mathfrak{p}_j\}=0$, 
$\{\mathfrak{q}_i,\mathfrak{p}_j\}=\delta_{ij}$, 
by means of relations
 \begin{eqnarray}\label{canqpinv}
&\mathfrak{q}_i=\frac{1}{B}\epsilon_{ij}(\mathcal{P}_j-\lambda\mathbb{P}_j)\,,\qquad
\mathfrak{p}_i=\frac{1}{2}(\mathcal{P}_i+\lambda\mathbb{P}_i)\,,\qquad
\lambda=\sqrt{ 1-\beta}\,.&
\end{eqnarray}
The non-commutative  vector variables are expressed then  in a  form 
\ba
&\mathcal{P}_i=\mathfrak{p}_i-\frac{1}{2}B\epsilon_{ij}\mathfrak{q}_j\,,\qquad
\lambda\mathbb{P}_i=\mathfrak{p}_i+\frac{1}{2}B\epsilon_{ij}\mathfrak{q}_j\,,\label{PpqBneq0}&\\
 &\mathbb{X}_i=\frac{1}{2}(1+\lambda^{-1}) \mathfrak{q}_i+ 
 \frac{1}{B}(1-\lambda^{-1})\epsilon_{ij} \mathfrak{p}_j\,,\qquad
 \mathbb{Y}_i=\frac{1}{2}(1+\lambda) \mathfrak{q}_i+ 
 \frac{1}{B}(1-\lambda)\epsilon_{ij} \mathfrak{p}_j\,.&\label{XYqpBneq0}
 \ea
The  canonical variables (\ref{canqpinv}) are defined so that 
$\mathcal{P}^i{}_{\vert_{B=0}}=\mathbb{P}^i{}_{\vert_{B=0}}=\mathfrak{p}^i$,
and $\mathbb{X}^i{}_{\vert_{B\rightarrow 0}}=\mathfrak{q}^i-\frac{1}{2}\theta\epsilon^{ij}
\mathfrak{p}_j$,  $\mathbb{Y}^i{}_{\vert_{B\rightarrow 0}}=\mathfrak{q}^i+\frac{1}{2}\theta\epsilon^{ij}
\mathfrak{p}_j$. Suchwise  the $\mathfrak{q}_i$  reduces at $B=0$ to the 
commutative coordinate $\mathcal{X}_i$ of the free particle in non-commutative plane,
see Eqs. (\ref{X+Ychi}), (\ref{HJ0m}).
On the other hand, one has $\mathbb{X}^i{}_{\vert_{\theta= 0}}=\mathbb{Y}^i{}_{\vert_{\theta= 0}}=
\mathfrak{q}^i$, $\mathcal{P}^i{}_{\vert_{\theta= 0}}=\mathfrak{p}^i-\frac{1}{2}B\epsilon^{ij}\mathfrak{q}_j$,
$\mathbb{P}^i{}_{\vert_{\theta= 0}}=\mathfrak{p}^i+\frac{1}{2}B\epsilon^{ij}\mathfrak{q}_j$,
that corresponds to the case 
of an ordinary Landau problem  in commutative plane.
Finally, the case $B=\theta=0$ corresponds to a free particle in 
commutative  plane:
$\mathbb{X}^i{}_{\vert_{B=\theta= 0}}=\mathbb{Y}^i{}_{\vert_{B=\theta= 0}}=
\mathfrak{q}^i$, $\mathcal{P}^i{}_{\vert_{B=\theta= 0}}=\mathbb{P}^i{}_{\vert_{B=\theta= 0}}=
\mathfrak{p}^i$. 
Integrals  $\mathbb{M}$  and 
$\breve{\mathbb{M}}$,  and  Hamiltonian (\ref{HmathP}) 
take the form
\ba\label{Mcalfrak}
&\mathbb{M}=\epsilon_{ij}\mathfrak{q}_i\mathfrak{p}_j:=\mathfrak{M},\qquad
\breve{\mathbb{M}}=\Omega^{-1}\mathfrak{H}_{\text{osc}}\,,\qquad
\mathfrak{H}_{\text{osc}}=\frac{1}{2m}\mathfrak{p}_i^2+\frac{1}{2}m\Omega^2\mathfrak{q}_i^2\,,&\\
\label{Hosciso}
&\mathbb{H}=\frac{1}{1-\beta}\left(\mathfrak{H}_{\text{osc}}
-\Omega\mathfrak{M}\right)\,,\qquad
\Omega=\frac{B}{2m}\,,&
\ea
where  $\mathfrak{H}_{\text{osc}}$ is a Hamiltonian of a planar isotropic harmonic oscillator. 
One finds
\ba
&\{\mathbb{M},\mathfrak{q}_i\}=\epsilon_{ij}\mathfrak{q}_j\,,\quad
\{\mathbb{M},\mathfrak{p}_i\}=\epsilon_{ij}\mathfrak{p}_j\,,\qquad
\{\breve{\mathbb{M}},\mathfrak{q}_i\}=-\frac{2}{B}\mathfrak{p}_i\,,\quad
\{\breve{\mathbb{M}},\mathfrak{p}_i\}=\frac{B}{2}\mathfrak{q}_i\,.
&
\ea
 
Define the 
``linearly polarized" creation-annihilation operators,
\begin{eqnarray}\label{ajqp}
&\hat{\mathfrak{a}}{}^\mp_j=\sqrt{\frac{m\vert \Omega\vert}{2\hbar}}\left(\hat{\mathfrak{q}}_j\pm i
\frac{1}{m\vert \Omega\vert}\,\hat{\mathfrak{p}}_j\right)\,,&
\end{eqnarray}
$[\hat{\mathfrak{a}}{}^-_j,\hat{\mathfrak{a}}{}^+_k]=\delta_{jk}$,
$[\hat{\mathfrak{a}}{}^+_j,\hat{\mathfrak{a}}{}^+_k]=[\hat{\mathfrak{a}}{}^-_j,\hat{\mathfrak{a}}{}^-_k]=0$,
which have a nature of the 2D vectors.
Here and in what follows we restore the explicit dependence on $\hbar$.
The ``ciricularly polarized" creation-annihilation operators
\ba\label{bpm+-}
&\hat{\mathfrak{b}}{}^-_\varepsilon=\frac{1}{\sqrt{2}}(\hat{\mathfrak{a}}{}^-_1
+i\varepsilon \hat{\mathfrak{a}}{}^-_2)\,,\qquad
\hat{\mathfrak{b}}{}^+_{\varepsilon}=
(\hat{\mathfrak{b}}{}^-_{\varepsilon})^\dagger\,,\qquad 
\varepsilon=\pm\,,&
\ea
satisfy the  commutation relations $[\hat{\mathfrak{b}}{}^-_\varepsilon,
\hat{\mathfrak{b}}{}^+_\varepsilon]=1$,
$[\hat{\mathfrak{b}}{}^\pm_-,\hat{\mathfrak{b}}{}^\pm_+]=0$.
From now on we identify the parameter $\varepsilon$ with the sign of magnetic field, 
$\varepsilon=\text{sgn}\,B\,.$ 
We have then  two pairs of circularly polarized operators, 
$\hat{\mathfrak{b}}^\pm_\varepsilon$ and $\hat{\mathfrak{b}}^\pm_{-\varepsilon}$, 
for magnetic field of a fixed sign.
The quantum analogs of 
the angular momentum,  integral $\breve{\mathbb{M}}$,  and Hamiltonian  
  take
 the form
\ba
&\hat{\mathbb{M}}=\hat{\mathfrak{M}}=\varepsilon \hbar (\hat{\mathfrak{N}}_{-\varepsilon}-\hat{\mathfrak{N}}_{\varepsilon})\,,\quad
\hat{\breve{\mathbb{M}}}=\varepsilon \hbar(\hat{\mathfrak{N}}_\varepsilon +
\hat{\mathfrak{N}}_{-\varepsilon} +1)\,,\quad
 \hat{\mathbb{H}}=\hbar \frac{\vert \Omega\vert}{1-\beta} (2\hat{\mathfrak{N}}_{\varepsilon}+1)\,,&
\ea
where $\hat{\mathfrak{N}}_{\varepsilon}=\hat{\mathfrak{b}}{}^+_{\varepsilon}\hat{\mathfrak{b}}{}^-_{\varepsilon}$,
$\hat{\mathfrak{N}}_{-\varepsilon}=\hat{\mathfrak{b}}{}^+_{-\varepsilon}\hat{\mathfrak{b}}{}^-_{-\varepsilon}$. 
In this alternative way we  reproduce the results of  Sec. \ref{Non-crit}  for the subcritical  phase 
of the NCLP. 
\vskip0.1cm
For the supercritical (chiral) phase of the NCLP,   
$\beta>1$,
we define
the canonical variables $({\widetilde{\mathfrak{q}}}_i, \widetilde{\mathfrak{p}}_i)$, 
$\sigma=d\,\widetilde{\mathfrak{p}}_i\wedge d\,\widetilde{\mathfrak{q}}_i$,
by relations similar to (\ref{canqpinv}), 
\ba\label{qipisuper}
&\widetilde{\mathfrak{q}}_i=\frac{1}{B}\epsilon_{ij}(\mathcal{P}_j-\tilde{\lambda}\check{\mathbb{P}}_j)\,,\qquad
\widetilde{\mathfrak{p}}_i=\frac{1}{2}(\mathcal{P}_i+\tilde{\lambda}\check{\mathbb{P}}_i)\,,\qquad
\tilde{\lambda}=\sqrt{\beta-1}\,.&
\ea
These canonical variables 
are obtained
from those for the subcritical phase by changing $\lambda=\sqrt{1-\beta}\rightarrow 
\widetilde{\lambda}=\sqrt{\beta-1}$,  $\mathbb{P}_i\rightarrow \check{\mathbb{P}}_i$.
 Relations inverse to (\ref{qipisuper}) 
are 
\begin{eqnarray}\label{deftildeP}
&\mathcal{P}_i=\widetilde{\mathfrak{p}}_i-\frac{1}{2}B\epsilon_{ij}
\widetilde{\mathfrak{q}}_j\,,\qquad
\tilde{\lambda}\check{\mathbb{P}}_i=\widetilde{\mathfrak{p}}_i+
\frac{1}{2}B\epsilon_{ij}\widetilde{\mathfrak{q}}_j\,.\qquad
\label{PpqBneq0}&
\end{eqnarray}
According to Eqs. (\ref{MVbra}) and (\ref{qipisuper}),
the canonical variables $(\widetilde{\mathfrak{q}}_i, 
\widetilde{\mathfrak{p}}_i)$ 
are not the 2D vectors with respect to the 
angular momentum $\mathbb{M}$, that  
we emphasize by supplying their characters with a tilde.
They transform, however, like  2D vectors under the action of the integral
$\breve{\mathbb{M}}$.
In terms of  $(\widetilde{\mathfrak{q}}_i,\widetilde{\mathfrak{p}}_i)$,
we have
\ba\label{NCLPtilMM}
&\breve{\mathbb{M}}=\epsilon_{ij}\widetilde{\mathfrak{q}}_i\widetilde{\mathfrak{p}}_j:=
\breve{\mathfrak{M}}\,,\qquad
{\mathbb{M}}=\Omega^{-1}\widetilde{\mathfrak{H}}_{\text{osc}}\,,\qquad
\widetilde{\mathfrak{H}}_{\text{osc}}=\frac{1}{2m}\widetilde{\mathfrak{p}}_i^2+
\frac{1}{2}m\Omega^2\widetilde{\mathfrak{q}}_i^2\,,&\\
\label{Hosciso+}
&\mathbb{H}=\frac{1}{\beta-1}\left(\widetilde{\mathfrak{H}}_{\text{osc}}-\Omega
\breve{\mathfrak{M}}\right)\,,&
\ea
wherefrom  one finds 
\ba
&\{\breve{\mathbb{M}},\widetilde{\mathfrak{q}}_i\}=\epsilon_{ij}\widetilde{\mathfrak{q}}_j\,,\quad
\{\breve{\mathbb{M}},\widetilde{\mathfrak{p}}_i\}=\epsilon_{ij}\widetilde{\mathfrak{p}}_j\,,\qquad
\{{\mathbb{M}},\widetilde{\mathfrak{q}}_i\}=-\frac{2}{B}\widetilde{\mathfrak{p}}_i\,,\quad
\{{\mathbb{M}},\widetilde{\mathfrak{p}}_i\}=\frac{B}{2}\widetilde{\mathfrak{q}}_i\,.
&
\ea 
Analogously to (\ref{ajqp}) and (\ref{bpm+-}), one defines the ``linearly polarized", $\hat{\widetilde{\mathfrak{a}}}{}^\pm_j$,
and ``circularly polarized", $\hat{\widetilde{\mathfrak{b}}}{}^\pm_{\varepsilon}$, 
$\hat{\widetilde{\mathfrak{b}}}{}^\pm_{-\varepsilon}$,  creation-annihilation
operators  in terms of 
$\hat{\widetilde{\mathfrak{q}}}_i$ and $\hat{\widetilde{\mathfrak{p}}}_i$, and we obtain 
\ba
&\hat{\mathbb{M}}= \varepsilon \hbar(\hat{\widetilde{\mathfrak{N}}}_\varepsilon
+\hat{\widetilde{\mathfrak{N}}}_{-\varepsilon}+1)\,,\qquad
\hat{\breve{\mathfrak{M}}}=\varepsilon\hbar(\hat{\mathfrak{\widetilde{N}}}_{-\varepsilon}-\hat{\mathfrak{\widetilde{N}}}_\varepsilon)\,,\qquad
\hat{\mathbb{H}}=\hbar\frac{\vert\Omega\vert}{\beta-1} (2\hat{\mathfrak{\widetilde{N}}}_{\varepsilon}+1)\,,&
\ea
where $\hat{\widetilde{\mathcal{N}}}_{\varepsilon}=
\hat{\widetilde{\mathfrak{b}}}{}^+_{\varepsilon}\hat{\widetilde{\mathfrak{b}}}{}^-_{\varepsilon}$, 
$\hat{\widetilde{\mathcal{N}}}_{-\varepsilon}=
\hat{\widetilde{\mathfrak{b}}}{}^+_{-\varepsilon}\hat{\widetilde{\mathfrak{b}}}{}^-_{-\varepsilon}$.
In this way we reproduce the results for the supercritical (chiral) phase  of the  NCLP    
in the way alternative to that presented in the previous section.

\subsection{Conformal bridge transformation for the NCLP} 
Consider a quantum  free particle system in 2D Euclidean space,
which is described by the operators of canonical coordinates $\hat{q}_i$ and momenta
$\hat{p}_i$.  
Its generators of the conformal $\mathfrak{sl}(2,\R)$
symmetry  
are the 2D analogs of
those considered in Appendix 
with restored and 
explicitly shown dependence 
on parameters $m$, $\Omega=\frac{B}{2m}$    and constant $\hbar$,
\ba\label{hatJ2d}
&\hat{J}_0=\frac{1}{2\hbar\vert\Omega\vert}\hat{H}_+\,,\qquad
\hat{J}_1=-\frac{1}{2\hbar \vert\Omega\vert}\hat{H}_-\,,\qquad
\hat{J}_2=-\frac{1}{2\hbar}\hat{D}\,,&\\
\label{Hpm2D}
&\hat{H}_\pm=\frac{1}{2m}\hat{p}_i^2\pm 
\frac{1}{2}m\Omega^2\hat{q}_i^2\,,\qquad
\hat{D}=\frac{1}{2}(\hat{q}_i\hat{p}_i+
\hat{p}_i\hat{q}_i)\,.&
\ea
Operators $\hat{H}_+$ and $\hat{H}_-$ correspond to 
the Hamiltonians of the quantum 2D isotropic harmonic 
and inverted harmonic oscillators, respectively,
while $\hat{D}$ is the generator of dilatations.
A two-dimensional analog 
of the similarity transformation (\ref {SigmaS}) is
\ba\label{Stransfor}
&\hat{O}{}'=\hat{\mathfrak{S}}\hat{O}\hat{\mathfrak{S}}{}^{-1}\,,
\qquad
\hat{\mathfrak{S}}=\exp(-\frac{\pi}{2}\hat{J}_1)\,,&
\ea
where $\hat{J}_1$ is given by Eqs. (\ref{hatJ2d}) and (\ref{Hpm2D}).
Transformation (\ref{Stransfor})
yields a mapping
\ba
&\left( i\sqrt{\frac{2}{\vert B\vert \hbar}}\,\hat{p}_j ,\,\sqrt{\frac{\vert B\vert}{2\hbar}}\,\hat{q}_j,\, i\sqrt{\vert B\vert \hbar}\,\hat{a}^+_j,\,
\sqrt{\frac{\hbar}{\vert B\vert}}\,\hat{a}^-_j,\,
\hat{H}_+,\,\hat{H}_-,\, i\hat{D},\, \hat{M}
\right)\rightarrow&\nonumber\\
& \left(\hat{a}_j^-,\, \hat{a}_j^+,\, \hat{p}_j,\, \hat{q}_j,\, -i\hat{D},\, \hat{H}_-,\, \hat{H}_+,\, \hat{M}\right)\,,&\label{CBTlist}
\ea
where $\hat{a}^\pm_j$ are the ``linearly polarized" creation-annihilation operators of the form 
(\ref{ajqp}), and 
$\hat{M}=\epsilon_{ik}\hat{q}_j\hat{p}_k=i\hbar\epsilon_{jk}\hat{a}^-_j\hat{a}^+_k$.
Introduce the complex canonical coordinates and 
momenta,
\ba\label{wpwdef}
&w_\varepsilon=\frac{1}{\sqrt{2}}(q_1+i\varepsilon q_2)\,,\qquad
\bar{w}_\varepsilon=w_{-\varepsilon}\,,\qquad 
p_\varepsilon=\frac{1}{\sqrt{2}}(p_1-i\varepsilon p_2)\,,\qquad
\bar{p}_\varepsilon=p_{-\varepsilon}\,,\quad
&
\ea
with the nontrivial brackets $\{w_\varepsilon,p_\varepsilon\}=
\{\bar{w}_\varepsilon,\bar{p}_\varepsilon\}=1$.  According to  (\ref{Stransfor}), 
their quantum analogs 
 are transformed into  the ``circularly polarized" creation-annihilation operators,
\ba\label{wwbb}
&(\hat{w}_\varepsilon,\,\hat{p}_\varepsilon)\rightarrow \big(\sqrt{\frac{2\hbar}{\vert B\vert}}\,\hat{b}^+_{-\varepsilon},\,
-i\sqrt{\frac{\hbar\vert B\vert}{2}}\,\hat{b}^-_{-\varepsilon}\big)\,,\qquad
(\hat{\bar{w}}_\varepsilon,\,\hat{\bar{p}}_\varepsilon)\rightarrow \big(\sqrt{\frac{2\hbar}{\vert B\vert}}\,\hat{b}^+_{\varepsilon}
,\,
-i\sqrt{\frac{\hbar\vert B\vert}{2}}\,\hat{b}^-_{\varepsilon}\big)\,,\qquad
&
\ea
which are linear combinations of $\hat{a}^\pm_j$ analogous to (\ref{bpm+-}).
One also finds that 
\ba
&i\hat{D}=\frac{i}{2}(\hat{q}_j\hat{p}_j+\hat{p}_j\hat{q}_j)\,\,\rightarrow \,\,
\hat{\breve{{M}}}=\hbar(\hat{N}_\varepsilon+
\hat{N}_{-\varepsilon}+1)\,,\quad
&\\
&
\hat{M}=\epsilon_{ij}\hat{q}_i\hat{p}_j\,\,\rightarrow\,\, \hat{{M}}=
\varepsilon\hbar(\hat{N}_{-\varepsilon}-
\hat{N}_{\varepsilon})\,,\qquad 
\hat{N}_\varepsilon=\hat{b}^+_\varepsilon\hat{b}^-_\varepsilon\,,\quad
 \hat{N}_{-\varepsilon}=\hat{b}^+_{-\varepsilon}
\hat{b}^-_{-\varepsilon}\,.&
\ea

The  angular momentum operator 
$\hat{M}$ commutes with the conformal symmetry generators (\ref{hatJ2d}). 
In the coordinate representation one has  $i\hat{D}=\hbar(w_\varepsilon\partial_{w_\varepsilon}+
\bar{w}_\varepsilon\partial_{\bar{w}_\varepsilon}+1)$ and
$\hat{M}=\varepsilon \hbar(w_\epsilon\partial_{w_\varepsilon}-\bar{w}_\varepsilon
\partial_{\bar{w}_\varepsilon})$.
The set of functions 
\ba\label{Phinn}
\Phi_{n_+,n_-}=w_\varepsilon^{n_+}{\bar{w}}_\varepsilon^{n_-}\,,\qquad n_+,\,n_-=0,1,\ldots,
\ea
is a set of formal eigenstates of the Wick rotated  dilatation and momentum operators,
\ba
&i\hat{D}\Phi_{n_+,n_-}=\hbar(n_+ +n_- +1)\Phi_{n_+,n_-}\,,\qquad
\hat{M}\Phi_{n_+,n_-}=\hbar(n_+ - n_-)\Phi_{n_+,n_-}\,.&
\ea
As $\hat{H}_0=
\frac{1}{2m}\hat{p}_i{}^2=-\frac{1}{m}\partial_{w_\varepsilon}\partial_{\bar{w}_\varepsilon}$, 
and $[\hat{M},\hat{H}_0]=0$,
$[i\hat{D},\hat{H}_0]=-2\hbar\hat{H}_0$, 
these states also are Jordan states of the free particle Hamiltonian corresponding to zero 
energy \cite{CJP,CarPlyu},
$
(\hat{H}_0)^{n_+ + n_- + 1}\Phi_{n_+,n_-}=0\,.
$ 
The state $\Phi_{0,0}$ is annihilated by the operators $\hat{p}_{\varepsilon}$ and $\hat{\bar{p}}_{\varepsilon}$,
and in accordance with (\ref{phinosc}),
is transformed, up to a normalization,  into the 
vacuum state of the operators $\hat{b}^-_\varepsilon$
and  $\hat{b}^-_{-\varepsilon}$, 
$\hat{b}^-_\varepsilon\Phi'_{0,0}=\hat{b}^-_{-\varepsilon}\Phi'_{0,0}=0$, where 
$\Phi'_{0,0}=\hat{\mathfrak{S}}\Phi_{0,0}\propto \exp(-\frac{m\vert\Omega\vert}{\hbar}w_\varepsilon
\bar{w}_\varepsilon)$.
In correspondence with (\ref{phinosc}), in the Fock representation for the 
``circularly polarized"
creation-annihilation operators,  the  states 
$\ket{n_+,n_-}$
 are 
the common eigenstates of the number operators $\hat{N}_\varepsilon$ and
$\hat{N}_{-\varepsilon}$. In the holomorphic representation these are, up to a normalization, 
the transformed states  (\ref{Phinn}), 
$\Phi'_{n_+,n_-}=\hat{\mathfrak{S}}\Phi_{n_+,n_-} $,  
cf. (\ref{phinosc}).

Identifying   now  the canonical variables (\ref{canqpinv}) with 
$(q_i, p_i)$, 
we generate the non-chiral (subcritical) phase of the NCLP   
from the 
free particle in a plane by applying to the latter 
the CBT 
(\ref{Stransfor}).   
The pre-images of the basic operators $\hat{\mathcal{P}}_i$ and 
$\hat{\mathbb{P}}_i$ 
 of the NCLP are the corresponding operators of the free particle, 
\ba\label{wwbb++}
&(\hat{w}_\varepsilon,\,\hat{p}_\varepsilon)\rightarrow 
\big(
-\frac{i}{\vert B\vert}\frac{1}{\sqrt{2}}(\hat{\mathcal{P}}_1+i\varepsilon\hat{\mathcal{P}}_2),\,
\frac{1}{\sqrt{2}}(\hat{\mathcal{P}}_1-i
\varepsilon\hat{\mathcal{P}}_2)\big)\,,
\qquad
&\\
\label{wwbb+++}
&
(\hat{\bar{w}}_\varepsilon,\,\hat{\bar{p}}_\varepsilon)\rightarrow \big(
-\frac{i}{\vert B\vert}\frac{\lambda}{\sqrt{2}}(\hat{\mathbb{P}}_1-i\varepsilon\hat{\mathbb{P}}_2),\,
\frac{\lambda}{\sqrt{2}}(\hat{\mathbb{P}}_1+i
\varepsilon\hat{\mathbb{P}}_2)\big)\,.
\qquad
&
\ea
The pre-images of the quadratic  integrals are identified from the CBT map
\ba\label{MtilMiDH}
&\hat{M}\rightarrow \hat{\mathbb{M}}\,,\qquad
i \hat{D}\rightarrow \hat{\breve{\mathbb{M}}}\,,\qquad
\frac{\vert\Omega\vert}{1-\beta}(i\hat{D}-\varepsilon\hat{M})\rightarrow \hat{\mathbb{H}}\,.&
\ea
The linear combinations of  $\hat{\mathcal{P}}_j$ and 
$\hat{\mathbb{P}}_j$ appearing in (\ref{wwbb++}),
 (\ref{wwbb+++})
 are represented  in terms of the circularly polarized
 creation-annihilation operators according to 
 Eq.  (\ref{wwbb}).
The pre-images of  
$\hat{\mathbb{X}}_i$ and $\hat{\mathbb{Y}}_i$ are  found from (\ref{wwbb++}),
 (\ref{wwbb+++})
by using Eq. (\ref{XYrelPP}).

\vskip0.1cm

In the supercritical phase we identify the canonical variables 
$\widetilde{\mathfrak{q}}_i$ and $\widetilde{\mathfrak{p}}_i$
given by Eq. (\ref{qipisuper}) 
with the canonical variables $q_i$ and $p_i$ of a free particle.
Then the similarity transformation
(\ref{Stransfor}) yields us the correspondence 
of the form (\ref{wwbb}),   (\ref{wwbb++}), (\ref{wwbb+++}), 
 with 
 $\lambda$ and $\hat{\mathbb{P}}_i$ 
 changed    for
 $\widetilde{\lambda}$ and $\hat{\check{\mathbb{P}}}_i$. 
 The pre-images of the operators $\hat{\mathbb{X}}_i$ and $\hat{\mathbb{Y}}_i$
are found by using Eqs. (\ref{XYrelPP}) and  (\ref{widetildePdef}).
 The map (\ref{MtilMiDH}) is changed here for
 \ba\label{MtilMiDH+}
&\hat{M}\rightarrow \hat{\breve{\mathbb{M}}}\,,\qquad
i\varepsilon \hat{D}\rightarrow \hat{{\mathbb{M}}}\,,\qquad
\frac{\vert\Omega\vert}{\beta-1}(i\hat{D}-\varepsilon \hat{M})\rightarrow \hat{\mathbb{H}}\,.&
\ea
 In both sub- and super- critical phases the last relations in 
 (\ref{MtilMiDH}) and (\ref{MtilMiDH+})
 can be presented in a universal form 
 \ba
 &\frac{\vert\Omega\vert}{\vert 1- \beta\vert }(i\hat{D}-\varepsilon \hat{M})\rightarrow \hat{\mathbb{H}}\,.&
 \ea
While the angular momentum 
of the free particle  $\hat{M}$ transforms into the angular momentum of the NCLP   
in the subcritical  phase,
and 
the Wick rotated generator of dilatations multiplied by $\varepsilon$
transforms in  the integral $\hat{\mathbb{\breve{\mathbb{M}}}}$,   the 
CBT from the free particle into  the chiral phase of the NCLP transmutes 
 $\hat{M}$ and $i\varepsilon \hat{D}$ into the integrals 
$\hat{\breve{\mathbb{M}}}$ and $\hat{\mathbb{M}}$, respectively.

Since in the case of a free particle in non-commutative plane we have
$\mathcal{P}^i{}_{\vert_{B=0}}=\mathbb{P}^i{}_{\vert_{B=0}}=\mathfrak{p}^i$,
and $\mathfrak{q}^i$ reduces to  the coordinate $\mathcal{X}^i$ defined  in 
Eq. (\ref{X+Ychi}), the CBT   
can  also be reinterpreted 
as a non-unitary mapping from a free particle system  in non-commutative plane ($B=0$, $\theta\neq 0$) 
into 
the NCLP  ($B\neq 0$, $\theta\neq 0$) in non-chiral and chiral phases 
by changing the coordinate  variable $q_i$ of the free particle   for $ \mathcal{X}_i$
 in the above relations.

 As the CBT is an invertible non-unitary transformation,
 one can also relate the non-chiral and chiral phases of the 
 NCLP by means of the  composition of the corresponding inverse CBT 
 from the subcritical phase into a ``virtual" 
 free particle with  the direct CBT from 
 the free particle into the supercritical phase. 
 In such a composition, it is convenient to take in both 
 phases the same value for magnetic field $B$,
 and choose the non-commutative parameter 
 $\theta$ in the first, inverse CBT,   such that 
 $\beta=B\theta<1$, and take  $\theta'$  so that
 $\beta'=B\theta' >1$ for the second, direct CBT
 from the ``virtual" free particle  in commutative (or non-commutative) plane
 into the chiral phase of the NCLP.

For completeness we just note that  the  critical phase $\beta=1$
can be generated by the  CBT from a 1D free particle, 
see Appendix. In this case the Wick rotated dilatation operator 
multiplied by $-\varepsilon_\theta=-\text{sgn}\,\theta$
is mapped into the angular momentum  $\hat{\mathbb{M}}$.

\subsection{Conformal bridge transformation for the two-vortex system} 
The CBT for the two-vortex system
can be realized based on the  correspondence with the NCLP
established in Sec. \ref{Non-crit}.
We restrict ourselves here by presenting the 
basic relations necessary for the construction,  
and making brief comments on  the transformation.

The correspondence (\ref{corr1}), (\ref{corr2}) between the NCLP and 
the two-vortex system
is extended by relations  for  the integral $\breve{\mathbb{M}}$  
and spatial reflection of $\mathbb{P}_i$,  
\ba\label{checkri}
\breve{\mathbb{M}}\sim \breve{M}_\Gamma=M_\Gamma+\varrho\, r_i^2\,,\qquad
\check{\mathbb{P}}_i=(-\mathbb{P}_1,\mathbb{P}_2)\sim \gamma_2\check{r}_i\,,\qquad
\check{r}_i=(r_2,r_1)\,.
\ea
We also  denote  
 the direct analog of the NCLP Hamiltonian,
cf. Eq. (\ref {H0G0}),
\ba\label{HadefHG}
& \mathcal{H}_\Gamma
:=\frac{1}{2}\gamma_2^2 r_i^2=\frac{1}{2}\gamma_2^2\exp(-2\kappa^{-1}H_\Gamma)\sim
\mathbb{H}\,.&
\ea

In the  two-vortex system with strenths $\gamma_a$ of the opposite sign, 
the canonical variables $(q_i,p_i)$  can be defined 
 based on the correspondence   
 with the NCLP and 
Eq. (\ref{canqpinv}),
\ba\label{qp<0}
&{q}_i=\frac{1}{\Gamma}\left((\gamma_1+\lambda\gamma_2)
x^i_1+\gamma_2(1-\lambda)x^i_2\right)\,,\qquad
{p}_i=\frac{1}{2}\epsilon_{ij}\left(\gamma_1-\lambda \gamma_2)x^j_1
+\gamma_2 (1+\lambda)x^j_2\right)\,,\qquad&
\ea
where $\lambda=\sqrt{-\gamma_1/\gamma_2}$. In the limit  $\Gamma\rightarrow 0$, 
one has ${q}^i_{{}\vert_{\Gamma\rightarrow 0}}=\frac{1}{2}(x^i_1+x^i_2)=\chi^i$,
${p}^i_{{}\vert_{\Gamma\rightarrow 0}}=\Pi^i$, where 
 $\chi^i$ and $\Pi^i$ are the canonical variables of the vortex-antivortex system,
 see Eq. (\ref{G0Pi}).
 Inverting  (\ref{qp<0}),  one finds 
\ba
&x^i_1=\frac{1}{2}(1+\lambda^{-1}){q}_i-\frac{1}{\Gamma}(1-\lambda^{-1})\epsilon_{ij}{p}_j\,,
\qquad
x^i_2=\frac{1}{2}(1+\lambda){q}_i-\frac{1}{\Gamma}(1-\lambda)\epsilon_{ij}{p}_j\,,&\\ 
&P_i=p_i+\frac{1}{2}\Gamma\epsilon_{ij}q_j\,,\qquad
\lambda\gamma_2\, r_i=\epsilon_{ij}p_j+\frac{1}{2}\Gamma q_i\,.&
\ea
In terms of  canonical variables (\ref{qp<0}), 
the integrals $M_\Gamma$,  $\breve{M}_\Gamma$ and $\mathcal{H}_\Gamma$  are 
presented in a form similar to (\ref{Mcalfrak}), (\ref{Hosciso}), 
\ba
&M_\Gamma=\epsilon_{ij}q^ip^j:=\mathcal{M}\,,\qquad
\breve{M}_\Gamma=\Omega^{-1}\mathcal{H}_{\text{osc}}\,,
\qquad 
\mathcal{H}_{\text{osc}}=\frac{1}{2}p_i^2+\frac{1}{2}\Omega^2q_i^2\,,
&\\
&
\mathcal{H}_\Gamma=
\lambda^{-2}(\mathcal{H}_{\text{osc}} - \Omega\mathcal{M})\,,\qquad
\Omega=-\frac{1}{2}\Gamma\,.& \label{OmGam}
\ea
The free particle dilatation generator
is presented here in the form 
\ba\label{Dqpxx}
&D=q_ip_i= \text{sgn}\,\gamma_2\,\sqrt{-\kappa}\,\epsilon_{ij}x^i_1x^j_2\,.&
\ea

In the two-vortex system with vorticities of the same sign, $\kappa>0$,
the canonical variables $(\widetilde{q}_i,\widetilde{p}_i)$ 
can be defined by analogy with  (\ref{qipisuper}), 
\ba\label{tildevort1}
&\widetilde{q}_i=-\frac{1}{\Gamma}\epsilon_{ij}(P_j-\gamma_2\widetilde{\lambda}\,\check{r}_j)\,,\qquad
\widetilde{p}{\,}^j=P_i+\gamma_2\widetilde{\lambda}\,\check{r}_i\,,&
\ea
where $\widetilde{\lambda}=\sqrt{\gamma_1/\gamma_2}$.  
For the integrals $\breve{M}_\Gamma$, $M_\Gamma$ and $\mathcal{H}_\Gamma$
we obtain here a representation similar to (\ref{NCLPtilMM}),
(\ref{Hosciso+}), 
\ba\label{breveM2vort}
&\breve{M}_\Gamma=\epsilon_{ij}\widetilde{q}{\,}^i\widetilde{p}{\,}^j:=\breve{\mathcal{M}}\,,\qquad
M_\Gamma=\Omega^{-1}\widetilde{\mathcal{H}}_{\text{osc}}\,,\qquad
\widetilde{\mathcal{H}}_{\text{osc}}=\frac{1}{2}\widetilde{p}_i{}^2+\frac{1}{2}\Omega^2\widetilde{q}_i{}^2\,,&\\ 
&\mathcal{H}_\Gamma=\tilde{\lambda}^{-2}(\widetilde{\mathcal{H}}_{\text{osc}}-\Omega\breve{\mathcal{M}})\,,&
\ea
where $\Omega$ is defined as in (\ref{OmGam}), $\Omega=-\frac{1}{2}\Gamma$. Relation (\ref {Dqpxx})
is changed for 
$
D=\widetilde{q}_i\widetilde{p}_i=\text{sgn}\,\gamma_2\,\frac{\sqrt{\kappa}}{\Gamma}\,
\epsilon_{ij}P_i\check{r}_j$.

Having the complete correspondence list for the dynamical variables 
and integrals in the NCLP and the two-vortex system in the  cases of
vorticities $\gamma_1$ and $\gamma_2$ of the opposite and equal  signs,
one can apply the CBT to a free particle to obtain 
the quantum system described by the auxiliary Hamiltonian $\hat{\mathcal{H}}_\Gamma$
with nonzero total vorticity $\Gamma$.
The spectrum of the two-vortex system with $\Gamma\neq 0$ 
is obtained then  in accordance with Eq. (\ref{HadefHG}) by means 
of the relation
$\hat{H}_\Gamma=-\frac{1}{2}\kappa\log (2\gamma_2^{-2}\hat{\mathcal{H}}_\Gamma).$ 

One can also  generate the 
two-vortex systems with $\kappa=\gamma_1\gamma_2<0$, $\Gamma\neq 0$,
  and $\kappa>0$ 
from the
vortex-antivortex system with $\Gamma=0$. In this case,
the free particle coordinates $q_i$ are identified with the coordinates
$\chi_i$ of the vortex-antivortex system, while
the canonical momenta $p_i$ of the former 
are identified with the variables $\Pi_i$ of the latter.
Also, by means of the composition of the corresponding 
inverse and direct conformal bridge transformations,
the two-vortex system with $\kappa<0$, $\Gamma\neq 0$,   and
 the system with $\kappa>0$ can be related 
via the ``virtual" free particle,
or via the vortex-antivortex system with $\Gamma=0$.

\section{Summary, discussion  and outlook}\label{Section6}
Let us first summarize and discuss  the obtained results.

\vskip0.1cm
\noindent $\bullet$  We have established a correspondence  between 
the non-commutative Landau problem and the system of  two point vortices
both at the classical and quantum levels.
This is done  for all three different cases  of the NCLP:
its sub-critical  (non-chiral),  
super-critical (chiral), and   a critical phases.
\vskip0.1cm

As both systems have phase spaces of the same dimension (equal four 
in  non-critical cases)
and their symplectic structures are similar, a priori 
one would immediately  expect  that there is a way to reinterpret one system in terms 
of  another. However, in general case similarity of symplectic structures is not sufficient 
to establish a correspondence as
classical systems 
with the same symplectic structure in phase space  of dimension $n\geq 4$
can still have different nature in the sense of integrability,  that prevents 
the  reinterpretation of one system  in terms of another.
On the other hand, sometimes  
distinct integrable systems with different symplectic structures 
can be related by means of a nontrivial mechanism of coupling constant metamorphosis,
which employs a non-canonical transformation that exchanges the 
roles of a coupling constant and  the energy in Hamiltonian systems while preserving
integrability, and 
includes additionally a correction of order $\hbar^2$ at the quantum level  \cite{HGDR}.
 Finally,  bi-Hamlitonian systems are described by different symplectic structures
 and different Hamiltonians.

\vskip0.1cm
The  two considered systems  not only have  
a similar symplectic structure, but both 
are superintegrable, and have the same symmetries being 
invariant under time and space translations, and  rotations.
Based also on the similarity of   their  dynamics, we introduced an imaginary 
mirror particle in the NCLP and put 
the vector coordinates of 
the particle, $\vec{\mathbb{X}}$,  and of its imaginary partner,
$\vec{\mathbb{Y}}$, 
in correspondence with coordinates of point vortices.
 In this way 
the total vorticity parameter
$\Gamma=\gamma_1+\gamma_2$ of the two-vortex system  was set
 in accordance with the magnetic field value   $B$, $\Gamma\sim -B$, in the Landau problem. 
As a consequence,  a strength  of one of the vortices 
is mapped to $-\theta^{-1}$, 
where  $\theta$ is the non-commutativity parameter 
in the NCLP.

\vskip0.1cm
\noindent $\bullet$ 
We have found that the 
i) non-chiral ($\kappa=\gamma_1\gamma_2<0$, $\Gamma\neq 0$), 
ii) chiral ($\kappa>0$),  
and iii) stationary 
(when one of the vortex strengths  $\gamma_a$ vanishes) 
cases of the two-vortex system
correspond to the
 i) sub- ($\beta=B\theta<1$), 
 ii) super- ($\beta>1$), 
 and  iii) critical ($\beta=1$) 
phases in the NCLP.  The 
angular momentum takes on  the values of both signs in their non-chiral
phase i), and of one sign in cases ii)   and iii). 
In spite of the indicated correspondence, 
the frequency of the classical circular motion in non-chiral and chiral phases of the system of two vortices
is energy-dependent in contrast with  the NCLP.  Another difference, associated with distinct form
of Hamiltonians (\ref{H2vort}) and (\ref{HMXY}),   is that 
the quantized energy levels in both non-critical phases of the NCLP
are  bounded from below, while in the two-vortex system with   $\kappa<0$, $\Gamma\neq 0$,  
and  $\kappa>0$ 
they are  bounded from below and above, respectively.

\vskip0.1cm
\noindent $\bullet$
 The vortex-antivortex system  ($\Gamma=0$)
with its rectilinear trajectories  
and hidden $(1+1)D$ ``isospace" Lorentz symmetry  corresponds to a free particle ($B=0$,
$\theta\neq 0$) 
in the non-commutative plane. 
In this case, essential difference in classical dynamics of the systems is that
with energy increasing  of a free particle in non-commutative plane,
its velocity increases as well as that of its imaginary partner,
together with the increase  of the distance between their parallel trajectories,
while   the velocities of the vortex and anti-vortex 
decrease when the vortex-anti-vortex energy and the distance between
 their parallel trajectories increase.

\vskip0.1cm
\noindent $\bullet$
The case of the ordinary  (commutative) Landau problem ($\theta=0$, $B\neq 0$) is also
covered by the  correspondence with the two-vortex system 
through a  limit procedure in which one of the vortex strengths tends  to infinity.
In this limit, the  coordinates of the vortices coincide, $x^i_1=x^i_2$, 
and their components commute like  this  happens with 
the coordinates of the particle and its imaginary partner in the 
ordinary Landau problem, $\mathbb{X}_i=\mathbb{Y}_i$, 
$\{\mathbb{X}_i,\mathbb{X}_j\}=0$.

\vskip0.1cm
\noindent $\bullet$ A non-trivial nature of the revealed  correspondence is 
manifested in the fact that
a simple permutation symmetry of the vortices generates
 a weak-strong coupling duality  in the NCLP, $\mathcal{I}:\,\mathbb{H}\rightarrow
 (1-\beta)^{2}\mathbb{H}$, where $\mathbb{H}$ is the NCLP  Hamiltonian.
 The particular chiral, $\beta=2$,  and non-chiral,  $\beta=0$, cases
 are stable under this 
 weak-strong duality, and correspond  to the cases of 
 the vortices of equal strengths, $\gamma_1=\gamma_2$, and to the vortex-antivortex system
 with zero total vorticity,
 $\Gamma=0$, or to the indicated  limit case to be similar to the ordinary  
 Landau problem with  $\theta=0$, $B\neq 0$.
 
 \vskip0.1cm 
 We introduced a linear combination $\breve{\mathbb{M}}$  of the 
 angular momentum integral $\mathbb{M}$  and Hamiltonian $\mathbb{H}$  
 in the NCLP,  which generates 
 rotations of the 
 vector integral of motion  $\mathcal{P}_i$ and  of 
 a spatially reflected non-commutative momentum $\check{\mathbb{P}}_i$, 
 and found the analogs of these objects in  the two-vortex system. 
 This allowed us  to identify   Darboux coordinates 
 (canonical coordinates and momenta) in phase space, which are transformed as
  2D vectors relative either to the angular momentum 
 $\mathbb{M}$ or the modified integral 
 $\breve{\mathbb{M}}$
  in the non-chiral or chiral phases, respectively.
 
  \vskip0.1cm
Let us emphasize  that the CBT construction 
for  the NCLP in the chiral phase  ($\beta>1$) (and in  the system
of the two point vortices with the strengths $\gamma_a$ of the same sign)
requires to introduce into consideration the two indicated additional  objects  
$\breve{\mathbb{M}}$ and   $\check{\mathbb{P}}_i$. This 
is necessary  
because the CBT implies  
the construction of the canonical coordinates and momenta 
in the phase space of the NCLP (and the two-vortex system)
with their subsequent  replacement with  those of a free particle.
By Darboux theorem,
there is no problem to construct such variables in the 
phase space of  the NCLP or the two-vortex system
(see, for instance, ref. \cite{non-Lan2+} where this  was done  
for the NCLP  by applying certain Bogolyubov transformations).
However, the corresponding sets of canonical coordinates
and momenta constructed as  linear combinations 
of the NCLP vectors  $\vec{\mathbb{X}}$ and  $\vec{\mathbb{P}}$ 
(or of the vortex coordinates $\vec{x}_a$, $a=1,2$)
cannot have  a vector nature 
with respect to the angular momentum integral $\mathbb{M}$
($M_\Gamma$) in the chiral phase,  where the latter
takes on the values of only one sign in contrast  to  the 
angular momentum of a free particle.
On the other hand, the momentum integral $\mathcal{P}_i$ for the NCLP 
and the Poisson-commuting with it  
$\check{\mathbb{P}}_i$ are  transformed by 
$\breve{\mathbb{M}}$ as  the two-component vector objects.
The canonical coordinates and  momenta on the phase space
are constructed as linear combinations 
of $\mathcal{P}_i$ and $\check{\mathbb{P}}_i$, 
and in their terms  the integral $\breve{\mathbb{M}}$ 
takes  the  form of  the angular momentum of
a  2D free particle.  See Eqs. (\ref{qipisuper}) and (\ref{NCLPtilMM}), 
and Eqs. (\ref{tildevort1}), (\ref{breveM2vort}) for the corresponding chiral phase of the 
two-vortex system~\footnote{
The authors of  ref. \cite{non-Lan2+}  identify  the three phases of 
the NCLP, but focusing on the aspects of noncommutative quantum mechanics 
distinct  from those that we study here, 
their  sets of canonical coordinates and momentum variables 
are not of a vector nature both in the $\beta>1$ phase and in the $\beta<1$.}.

 \vskip0.1cm
\noindent $\bullet$
 Based on  the  two sets 
 of  canonical variables constructed separately for non-chiral and chiral phases,  
 we  built two forms of the conformal bridge 
 transformation  (CBT), which is  a non-unitary similarity transformation
that maps a  2D quantum free particle system
 into  non-chiral and chiral phases of the
 two-vortex and NCLP systems. 
 
 \vskip0.1cm
 \noindent $\bullet$ 
The generator of the CBT has a form of the evolution operator for
  the 2D inverted harmonic oscillator taken for a particular complex value 
  of the time  parameter $t =i\pi/4$. 
 It acts as the eighth-order root of  an identity  transformation in the phase space
  of  a 2D free particle, which changes the topological nature of its  
  $\mathfrak{sl}(2,\R)$ conformal symmetry  generators,
see Eq. (\ref{CBTlist}) and Fig. \ref{FigureCBT} in Appendix. 
 As a result, it produces  the common eigenstates of the Hamiltonian 
 and angular momentum 
 of the quantum two-vortex and NCLP systems both in chiral and non-chiral  phases
 from the simple monomial
  eigenfunctions of the angular momentum operator, which simultaneously 
  are formal  eigenfunctions  of the Wick rotated dilatation generator 
 of the 2D free particle. Such monomial  in coordinates  functions are simultaneously Jordan states corresponding 
 to zero energy of the free particle and  annihilated by its Hamiltonian 
 $\hat{H}_0$, or by  higher powers of  $\hat{H}_0$ \cite{CJP,CarPlyu}.

   \vskip0.1cm
  It is  interesting to note here some  formal 
  similarity of the CBT  in a simpler case 
  of the 2D isotropic harmonic  oscillator  \cite{CBT1}, where the Wick rotated dilatation generator 
  is transformed in  Hamiltonian, 
   with the reverse  picture  corresponding to the  radial quantization of the closed string  \cite{CFTbook}. 
  In string theory, after the Wick rotation of the evolution parameter on the string surface and the subsequent 
  mapping of the cylinder onto the complex plane,  by means  of an  exponential conformal transformation, 
  the generator of  time translations on the cylinder is transformed into a dilatation 
  generator, which takes on the  role of the Hamiltonian.

 \vskip0.2cm
\noindent $\bullet$ 
Here, unlike the  case of the 2D isotropic harmonic oscillator, 
a  linear combination of the Wick rotated dilatation 
 generator and angular momentum operator of 
the  free particle is transformed under the CBT 
 into the Hamiltonian operators of the two quantum systems 
 in corresponding  non-chiral or chiral phases.
 In the case of the mapping by the CBT into the non-chiral phase, 
 the angular momentum operator of the free particle 
 is  transformed into the angular momentum integral of the 
 two-vortex or  the NCLP systems, while the  Wick rotated 
 dilatation generator of the free particle  is mapped 
 into the integral $\hat{\breve{\mathbb{M}}}$ of the NCLP,
 or its analog in the two-vortex system. 
 In the case of the transformation into the chiral phase of the two systems,
 the Wick rotated dilatation generator of the free particle 
 transmutes into the angular momentum operator, while 
 its angular momentum transmutes  into the integral 
$\hat{\breve{\mathbb{M}}}$ of the NCLP and
its analog in the two-vortex system.

Such a phenomenon of transmutation
 into and from the  angular momentum, which  occurs  in the chiral phases
  of the two systems,    
  has not previously been observed in any of other  systems \cite{CBT1}--\cite{CBT4},
  \cite{CBT5}--\cite{CBT7}, 
   to which the CBT has been applied.

 \vskip0.2cm
 Thus, the CBT highlights an essential difference between 
the non-chiral and chiral phases in both considered systems from 
a new  perspective. 

\vskip0.2cm
\noindent $\bullet$
The nature of 
the chiral and non-chiral phases in both
quantum systems can be  related  to 
the   outer $Z_2$ automorphism of the conformal 
$\mathfrak{sl}(2,\R)$ algebra, 
see Appendix.

\vskip0.2cm
\noindent $\bullet$
 The CBT can also be applied to 
 map the vortex-antivortex system ($\Gamma=0$)
 and free particle in  the non-commutative plane ($B=0$, $\theta\neq 0$)
 into non-chiral and chiral phases of the
 two-vortex and NCLP systems, respectively.
 \vskip0.2cm
\noindent $\bullet$
Though the chiral and non-chiral phases of the systems 
are characterized by essentially different properties,
they can be mutually mapped by a composition of the  inverse and direct 
 conformal bridge transformations.

 \vskip0.2cm
The described picture of the mapping which transforms a linear combination 
 of the Wick rotated dilatation generator and the angular momentum of the "source system" 
   into the Hamiltonians of the "target systems" 
    together with the mapping of the corresponding eigenstates of the symmetry generators
     represents  some  development of the idea of different forms of dynamics of Dirac  \cite{Dirac}
     by incorporation into the construction of a kind of the Dyson map associated with  the $\mathcal{PT}$ 
symmetry \cite{Bender,CBT5,CBT6}.

\vskip0.2cm
 We expect that the established correspondences  and mappings 
 can be useful  for the theory of anyons and fractional Hall effect,
 where the point vortices and non-commutative quantum mechanics 
 play an important role.
 In this context it is interesting to mention  a recent application 
 of the point vortices in  the  rapidly developing topic 
of fracton phases 
in condensed matter physics, characterized  by
local excitations with restricted mobility
\cite{Fractons}.   
In particular, the single-vortex system (corresponding, as we showed, to critical phase in 
the NCLP) 
is identified there as  an immobile fracton.  The 
vortex-antivortex system  ($\Gamma=0$)   
is treated there as a ``lineon",
and a system of the two point vortices of the 
vorticity strengths of the same sign 
is considered in fractonic context as a 2-dipole.
Because of the established correspondence of the two-vortex systems
with different   phases of the NCLP, the latters  also can be related 
to simple fractonic systems.

 \vskip0.1cm
 We investigated the relationships between the classical and quantum 
 dynamics of the  planar two-vortex system, the noncommutative Landau problem,  and the free particle 
 in the commutative and noncommutative planes.
It seems to be  interesting  to generalize the obtained results for the case 
of the spherical and hyperbolic geometries \cite{non-Lan2+,ComHou,Comtet}.
The case of the hyperbolic geometry is of a particular interest
 in the light of the  non-relativistic conformal-invariant Schwartzian mechanics associated 
  with the low energy limit of the Sachdev-Ye-Kitaev model
  \cite{MalSta,MalStaYan}, 
  which  reveals a close relationship with 
  the particle dynamics on AdS${}_2$ 
  and Landau problem  on hyperbolic   plane \cite{MerTurVer,GanPol}.

\vskip0.3cm
\noindent
{\bf Acknowledgments}
\vskip0.2cm
\noindent
The work was partially supported by the FONDECYT Project 1190842 and the DICYT
Project 042331PS${}_{-}$Ayudante. 
We thank J. Gamboa and L. Inzunza  for useful discussions.

 \appendix

\section*{Appendix}
\section{Conformal Bridge Transformation}

Consider a free particle in $d=1$  dimension, and  for the sake of simplicity set 
mass parameter
$m=1$.  The coordinate $q$ and canonical momentum $p$ can be 
combined into a two-component object  $\xi_\alpha^T=(q,p)$, $\{\xi_\alpha,\xi_\beta\}=\epsilon_{\alpha\beta}$.
Quadratic in $\xi_\alpha$ functions 
$H_0=\frac{1}{2}p^2$, $D=qp$ and  $K=\frac{1}{2}q^2$ generate the $\mathfrak{sl}(2,\R)$ algebra
of the form (\ref{DKH0}).  
A linear combination of $H_0$ and $K$,
$H_+=\frac{1}{2}(p^2+q^2):=2J_0$ is a compact generator of  the $\mathfrak{sl}(2,\R)$, 
and  can be considered as the Hamiltonian of the 
harmonic oscillator of frequency $\omega=1$. Another linear combination
$H_-=\frac{1}{2}(p^2-q^2):=-2J_1$,  which is a non-compact $\mathfrak{sl}(2,\R)$  generator,  
corresponds to the Hamiltonian of the inverted oscillator. Identifying  $J_2=-\frac{1}{2}D$,
one finds that in accordance with Eq. (\ref{sl2(1)}), the  $\mathfrak{sl}(2,\R)$  generators
$J_0$, $J_1=\frac{1}{2}(J_++J_-)$ and $J_2=\frac{i}{2}(J_--J_+)$
satisfy the algebra
$\{J_0,J_j\}=\epsilon_{jk}J_k$, $j,k=1,2,$
$\{J_j,J_k\}=-\epsilon_{jk}J_0$.
This can be unified and presented in a compact (2+1)D form
\ba\label{JmuJnu}
&\{J_\mu,J_\nu\}=-\epsilon_{\mu\nu\lambda}J^\lambda\,,&
\ea
where $\mu,\nu,\lambda=0,1,2$,  $\epsilon_{\mu\nu\lambda}$ is an antisymmetric tensor,
$\epsilon^{012}=1$, and the metric tensor is taken to be 
$\eta_{\mu\nu}=\text{diag}\,(-1,1,1)$.

A phase space function $F(\xi_\alpha)$  generates canonical transformations
\ba\label{Ftrans}
&F\,:\,\, \xi_\alpha\rightarrow \xi'_\alpha=\exp(\tau F)\star\xi_\alpha=
\xi_\alpha +\sum_{n=1}^\infty \frac{\tau^n}{n!}\{F,\{\ldots,\{F,\xi_\alpha\underbrace{\}\ldots\}}_{{n}}.&
\ea
In this way  $H_+$ with  $\tau\in \R$ generates rotations of $\xi_\alpha$,
\ba\label{H+trans}
&H_+\,:\,\, \xi_\alpha\rightarrow \xi'_\alpha=R_{\alpha\beta}\xi_\beta\,,\qquad
R_{\alpha\beta}=
\left(
\begin{array}{cc}
 \cos\tau &  -\sin\tau \\
  \sin\tau   &  \cos\tau 
\end{array}
\right).&
\ea
Analogously one has
\ba\label{2Dtrans}
&D\,:\,\, q\rightarrow q'=e^{-\tau}q\,,\qquad
p\rightarrow p'=e^\tau p\,,&\\ 
\label{H-trans}
&H_-\,:\,\, \xi_\alpha\rightarrow \xi'_\alpha=L_{\alpha\beta}\xi_\beta\,,\qquad
L_{\alpha\beta}=
\left(
\begin{array}{cc}
 \cosh\tau &  -\sinh\tau \\
 - \sinh\tau   &  \cosh\tau 
\end{array}
\right).&
\ea
So $H_-$ generates hyperbolic rotations of $\xi_\alpha$~\footnote{
The $\mathfrak{sl}(2,\R)$ generators $J_0$ and  $J_j$ 
 produce transformations of the form 
 (\ref{H+trans}) and  (\ref{2Dtrans}), (\ref{H-trans}) 
 with $\tau$ changed for $\tau/2$ and $-\tau/2$, respectively, which means that 
 $\xi_\alpha$ is an $\mathfrak{so}(2,1)\cong \mathfrak{sl}(2,\R)$
 spinor \cite{MPAnn}.}.
 The action of $H_+$ 
on the classical analogs of the creation-annihilation
operators,
$
a^\mp=\frac{1}{\sqrt{2}}(q\pm ip),
$
produces a  phase transformation,
$
H_+\,:\,\, a^\pm\rightarrow a'^\pm=e^{\pm i\tau}a^\pm\,,
$ 
 while the action of $D$  generates a Bogolyubov transformation,
$
D\,:\,\,a^-\rightarrow {a'^-}=a^-\cosh\tau - a^+ \sinh\tau ,$
$a^+\rightarrow {a'^+}=a^+\cosh\tau -  a^-\sinh\tau$.
A transformation of the form (\ref{H+trans}) with
$\tau=\pi/4$ yields 
a canonical pair 
$q'=\frac{1}{\sqrt{2}}(q-p)$,  
$p'=\frac{1}{\sqrt{2}}(q+p)$,  
and $a'{\,}^\mp=e^{\pm i\pi/4}a^\mp$.
Under the canonical transformation $(q,p)\rightarrow (q', p')$,
$H_+$ is invariant, $H_+\rightarrow H'_+=H_+$, but  
$D\rightarrow D'=-H_-$ and $H_-\rightarrow H'_-=D$.
In variables $q'$, $p'$  generator $D$  takes a form of the Hamiltonian of the 
inverted harmonic oscillator.
With taking into account relations (\ref{H-trans}), this 
is behind the emergent $(1+1)D$ Lorentz transformations 
(\ref{Lorentz}).  
Bearing in mind that $J_1=-\frac{1}{2}H_-$, $J_2=-\frac{1}{2}D$, 
one sees that  transformation  $(q,p)\rightarrow (q', p')$
corresponds to a rotation by $\pi/2$ of the non-compact 
pair of the $\mathfrak{sl}(2,\R)$  generators,
 $J_i\rightarrow J'_i=\epsilon_{ij}J_j$, $J_0\rightarrow J_0$.

The considered   transformations with real values of the parameter $\tau$
do not change a compact or non-compact nature of the $\mathfrak{sl}(2,\R)$
generators. 
Allowing  the parameter $\tau$ to take complex values, we obtain 
complex canonical transformations. 
Setting  $\tau=i\pi/4$, we find then 
that $H_-=-2J_1$ produces the  transformation
$\exp(-i\frac{\pi}{2}J_1)\star
(J_0,J_1,J_2)	=
(J'_0, J'_1, J'_2)=(iJ_2,J_1,iJ_0)$, 
 or, equivalently, 
 \ba\label{Dstar}
&\exp(i\frac{\pi}{4} H_-)\star
(H_+,H_-,iD)=
( -iD,H_-,H_+)\,.&
\ea
In this way the Wick rotated non-compact  $\mathfrak{sl}(2,\R)$ generator $D$
is transformed into the   compact generator $2J_0=H_+$.
At the level of the phase space coordinates $q$, $p$ and their complex 
linear combinations $a^\pm$ we have
$\exp(i\frac{\pi}{4} H_-)\star
(q,p,a^-,a^+)=(a^+,-ia^-,q,-ip)$, 
that is  illustrated by  
Figure \ref{FigureCBT}.
From the viewpoint of the linear phase space variables,
this complex canonical transformation is of a nature of 
the eighth-order root  of the identity transformation.
For the $\mathfrak{sl}(2,\R)$ algebra, this acts as 
the fourth-order  root of  the identity  transformation. 
\vskip-0.5cm
\begin{figure}[H] 
\begin{center}
\includegraphics[scale=0.6]{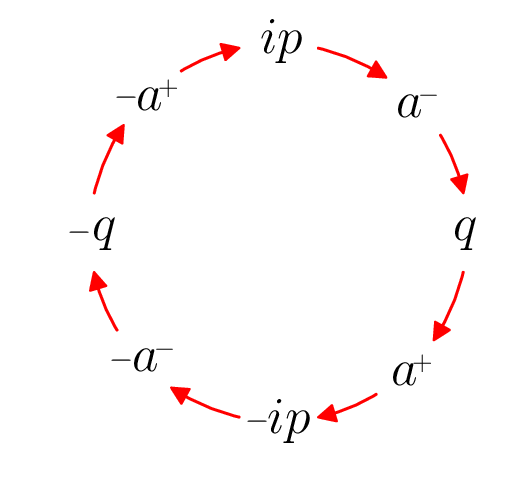}
\vskip-0.2cm
\caption{\small{Action of the conformal bridge transformation.} }
\label{FigureCBT}
\end{center}
\end{figure}

\vskip-0.5cm
Quantum analog of  (\ref{Ftrans})  with $i\tau\hat{F}$ changed for 
$\frac{\pi}{4}\hat{H}_-$
is  a similarity transformation
\ba\label{SigmaS}
& \hat{O}\rightarrow  \hat{O}{}'=\hat{\mathfrak{S}}\hat{O}\hat{\mathfrak{S}}{}^{-1}\,,
\qquad
\hat{\mathfrak{S}}=\exp(\frac{\pi}{4}\hat{H}_-)\,.&
\ea
The generator $\hat{\mathfrak{S}}$ of the 
similarity transformation (\ref{SigmaS}) 
has a form of the evolution operator for complex time $t=i\pi/4$
 of the inverted harmonic oscillator. 
This is a non-unitary,  non-local  transformation, which 
changes the anti-hermitian operator
$i\hat{D}=\frac{i}{2}(\hat{q}\hat{p}+ \hat{p}\hat{q})=
\frac{i}{2}(\hat{p}'{\,}^2-\hat{q}'{\,}^2)=i\hat{H}'_-$
into Hermitian  Hamiltonian operator $\hat{H}_+=\frac{1}{2}(\hat{p}{}^2+\hat{q}{}^2)$ 
of the quantum  harmonic oscillator,
and corresponds to the Dyson map, to which
the $\mathcal{PT}$ symmetry is  intimately related, see refs.  \cite{CBT5,CBT6}.
At the level of the canonical operators $\hat{q}$ and $\hat{p}$,  this
can be compared with 
transformation from the 
Schr\"odinger representation of the Heisenberg algebra 
to its Fock-Bargmann representation
\cite{CBT1}. 
 The monomials  $\phi_n=x^n$, $n=0,1,\ldots$,
 are the formal  eigenfunctions of the Wick rotated dilatation operator, 
 $i\hat{D}\phi_n=(n+\frac{1}{2})\phi_n$. 
 At the same time, $\phi_n$ are the Jordan states \cite{CJP}
 of the free particle corresponding to its zero energy eigenvalue,
 $(\hat{H}_0)^{\lfloor\frac{n}{2}\rfloor}\phi_n=0$, where $\lfloor\frac{n}{2}\rfloor$ 
 is the integer part of $n/2$. 
 As a consequence, 
 the CBT applied to $\phi_n$ transforms them, up to a normalization, 
 into eigenstates of the quantum harmonic oscillator,
 \be\label{phinosc}
 \hat{\mathfrak{S}}\phi_n(q)\propto \psi_n(q)=\frac{1}{2^n\pi^{1/2}n!}H_n(q)e^{-q^2/2}\,,
 \ee
 where $H_n(q)$ are the Hermite polynomials.
 At the same time, the free particle plane wave eigenstates 
 and the Gaussian packet composed from them are transformed
 into coherent states and single-mode squeezed coherent states 
 of the quantum harmonic oscillator respectively, see  ref.  \cite{CBT1}.
 
 The described CBT can be generalized directly for the
 case of $d>1$ dimensions by supplying the phase space coordinates 
 $q$ and $p$
 with index $j=1,\ldots, d$, and defining $H_\pm=\frac{1}{2}(p_j^2\pm q_j^2)$,
 $D=q_jp_j$. In this way we obtain  generators of the 
 $\mathfrak{sl}(2,\R)$ algebra, which are  invariant under the $so(d)$ rotations. 
In the  case of $d=2$ dimensions one gets 
$\exp(i\frac{\pi}{4}H_-)\star (iD+\mu M)=H_++\mu M$,  
where $M=\epsilon_{ij}q_ip_j$ is the angular momentum
and $\mu$ is a constant.

 The $\mathfrak{sl}(2,\R)$  algebra (\ref{JmuJnu}) has an outer   $\Z_2$ automorphism
\ba\label{sl2Rauto}
&J_\mu\rightarrow J'_\mu\,,\qquad
(J'_0,J'_1,J'_2)=(-J_0,J_2,J_1)\,,&
\ea
related to the existence of the  infinite-dimensional unitary 
representations in which eigenvalues of
the compact generator $\hat{J}_0$ are bounded  from below 
or from above \cite{MPAnn}. 
Note that (\ref{sl2Rauto}) represents a kind
of the $\mathcal{PT}$-inversion applied to $J_\mu$,  which is 
 a composition of the 
time, $J_0\rightarrow -J_0$, and space, $J_i=(J_1,J_2)\rightarrow -\epsilon_{ij}\check{J}_j=(J_2,J_1)$,
cf. (\ref{checkri}), inversions acting in the (2+1)D space with coordinates $J_\mu$. 
If one takes  two copies of the 
 $\mathfrak{sl}(2,\R)$ algebra generated by 
 $J_\mu^{(a)}$, $a=1,2$,  $\{J_\mu^{(1)},J_\nu^{(2)}\}=0$, 
 the three sets of the  $\mathfrak{sl}(2,\R)$ generators 
 can be constructed, 
 $
 \mathcal{J}_\mu=J_\mu^{(1)}+J_\mu^{(2)},
 $  $ {\mathcal{J}}'_\mu=J'_\mu{}^{(1)}+J'_\mu{}^{(2)}$, 
 $\widetilde{\mathcal{J}}_\mu={J}_\mu^{(1)}+J'_\mu{}^{(2)}$,
 where $J'_\mu{}^{(a)}$ are 
 obtained from $J_\mu^{(a)}$ 
  by applying the $Z_2$ automorphism (\ref{sl2Rauto}). 
  If $J^{(a)}_\mu$ are realized in 
  terms of the  one-dimensional canonical variables
  $(q^{(a)}, p^{(a)})$, in particular, 
    $J^{(a)}_0=\frac{1}{2}H_+^{(a)}$,
    one finds that $ \mathcal{J}_0$ (${\mathcal{J}}'_0$)
    takes non-negative (non-positive) values, while 
    $\widetilde{\mathcal{J}}_0$ takes values on  the entire real line.
    At the quantum level,   $\widetilde{\mathcal{J}}_0$ 
     acts in the space which  corresponds to  the  direct sum of the 
    two infinite-dimensional (reducible) unitary  representations of the $\mathfrak{sl}(2,\R)$  algebra 
    mentioned above, each of which is, in turn,  a direct sum of the two irreducible representations
    with eigenvalues of the corresponding compact generators  
    shifted mutually in $1/2$.  In these representations,
      the Casimirs take the same 
    value, $\hat{J}^{(1)}_\mu \hat{J}^{(1)}{}^\mu=\hat{J}'^{(2)}_\mu \hat{J}'^{(2)}{}^\mu=
    -\alpha(\alpha-1)=3/16$,
    while  $\hat{J}^{(1)}_0$ and  $\hat{J}'^{(2)}_0$  take eigenvalues 
    $\alpha+n_1$ and   $-(\alpha+n_2)$
    with $\alpha=1/4,\, 3/4$, and $n_1, n_2=0,1,\ldots$.
    Here, the direct sums of the irreducible $\mathfrak{sl}(2,\R)$ 
     representations  with $\alpha=1/4$ 
    and $\alpha=1/4+1/2=3/4$ constitute  irreducible representations
    of the two copies of the $\mathfrak{osp}(2,1)$ superalgebra in which $\hat{J}^{(1)}_\mu$ and 
        $\hat{J}'^{(2)}_\mu$ are treated as even generators,
   while  $\hat{\xi}_\alpha^{(1)}$ and  $\hat{\xi}_\alpha^{(2)}$ 
  are  its  odd generators which mutually transform the corresponding  states
  from the  $\alpha=1/4$ and $\alpha=3/4$  subspaces \cite{MPAnn}.
    It is this picture related to the $Z_2$ automorphism 
    of the $\mathfrak{sl}(2,\R)$ algebra that underlies 
   different  properties of the angular momentum 
   in the sub- (non-chiral) and super- critical (chiral) phases of the NCLP and 
   in the corresponding two-vortex systems 
   with $\kappa=\gamma_1\gamma_2<0$, $\Gamma\neq 0$, 
   and $\kappa>0$,
   see the second relation in Eq. (\ref{HJclasN=2}).


\begin{thebibliography}{99}
 
 

\bibitem{ADS1}
J.~M.~Maldacena,
\emph{``The Large N limit of superconformal field theories and supergravity",}
\href{https://www.intlpress.com/site/pub/pages/journals/items/atmp/content/vols/0002/0002/a001/}{Adv.  Theor. Math. Phys. \textbf{2} (1998) 231}
\href{https://arxiv.org/abs/hep-th/9711200}{\textcolor{magenta}{[arXiv:hep-th/9711200]}}.

\bibitem{ADS2}
S. S. Gubser, I. R. Klebanov,  and A. M. Polyakov,
\emph{``Gauge Theory Correlators from Non-Critical String Theory",}
\href{https://www.sciencedirect.com/science/article/abs/pii/S0370269398003773?via%3Dihub}{Phys. Lett. B {\bf 428}  (1998) 105}
\href{https://arxiv.org/abs/hep-th/9802109}{\textcolor{magenta}{[arXiv:hep-th/9802109]}}.

\bibitem{ADS3}
E. Witten, \emph{``Anti De Sitter Space And Holography",}
\href{https://www.intlpress.com/site/pub/pages/journals/items/atmp/content/vols/0002/0002/a002/}{Adv. Theor. Math. Phys. {\bf 2}  (1998)  253}
\href{https://arxiv.org/abs/hep-th/9802150}{\textcolor{magenta}{[arXiv:hep-th/9802150]}}.

\bibitem{ADS4}
O. Aharony, S. S. Gubser, J. Maldacena,  H. Ooguri,  and Y. Oz,
\emph{``Large N Field Theories, String Theory and Gravity",}
\href{https://www.sciencedirect.com/science/article/abs/pii/S0370157399000836?via%3Dihub}{Phys. Rept. {\bf 323}  (2000) 183}
\href{https://arxiv.org/abs/hep-th/9905111}{\textcolor{magenta}{[arXiv:hep-th/9905111]}}.



\bibitem{ADS5}
C.~Leiva and M.~S.~Plyushchay,
\emph{``Conformal symmetry of relativistic and nonrelativistic systems and AdS/CFT correspondence",}
\href{https://www.sciencedirect.com/science/article/abs/pii/S0003491603001180}{Annals Phys. \textbf{307}  (2003) 372}
\href{https://arxiv.org/abs/hep-th/0301244}{\textcolor{magenta}{[arXiv:hep-th/0301244]}}.

\bibitem{ADS6}
D.~T.~Son,
\emph{``Toward an AdS$/$cold atoms correspondence: A geometric realization of the Schr\"odinger symmetry"},
\href{https://journals.aps.org/prd/abstract/10.1103/PhysRevD.78.046003}{Phys. Rev. D {\bf 78}  (2008) 046003}
\href{https://arxiv.org/abs/0804.3972}{\textcolor{magenta}{[arXiv:0804.3972 [hep-th]]}}.

\bibitem{ADS7}
K.~Balasubramanian and J.~McGreevy,
\emph{``Gravity Duals for Nonrelativistic CFTs"},
\href{https://journals.aps.org/prl/abstract/10.1103/PhysRevLett.101.061601}{Phys. Rev. Lett. {\bf 101}  (2008) 061601}
\href{https://arxiv.org/abs/0804.4053}{\textcolor{magenta}{[arXiv:0804.4053 [hep-th]]}}.

\bibitem{ADS8}
J.~L.~F. Barb\'on and C. ~A. ~Fuertes,
\emph{``On the spectrum of nonrelativistic $AdS/CFT$",}
\href{https://iopscience.iop.org/article/10.1088/1126-6708/2008/09/030}{JHEP \textbf{09}  (2008) 30}
\href{https://arxiv.org/abs/0806.3244}{\textcolor{magenta}{[arXiv:0806.3244 [hep-th]]}}.

\bibitem{ADS9}
C.~Chamon,~R.~Jackiw, S.~Y.~Pi, and L.~Santos,
\emph{``Conformal quantum mechanics as the $CFT_{1}$ dual
to $AdS_{2}$",} 
\href{https://www.sciencedirect.com/science/article/pii/S0370269311006447}{Phys. Lett. B. \textbf{701}  (2011) 503}
\href{https://arxiv.org/abs/1106.0726}{\textcolor{magenta}{[arXiv:1106.0726 [hep-th]]}}.

\bibitem{Arnold}
  V. I. Arnold and V. A. Vassiliev, \emph{``Newton's principia read 300 years later",} 
  \href{https://www.ams.org/journals/notices/198911/198911FullIssue.pdf}{ Not. Am. Math. Soc. {\bf 36}  (1989) 1148}; 
 \href{https://www.ams.org/journals/notices/199002/199002FullIssue.pdf}{{\bf 37}  (1990) 144 (addendum)}.
  
\bibitem{Hoj}
S.A. Hojman, S. Chayet, D. Nunez and M.A. Roque, 
 \emph{``An algorithm to relate general solutions of different bidimensional
problems",}  
\href{https://aip.scitation.org/doi/abs/10.1063/1.529255}{ J. Math. Phys. {\bf 32}  (1991) 1491}.

 
  \bibitem{InoJun}
  A. Inomata and G. Junker,
   \emph{``Power law duality in classical and quantum mechanics",} 
   \href{https://www.mdpi.com/2073-8994/13/3/409}{Symmetry  {\bf 13}  (2021) 409}
   \href{https://arxiv.org/abs/2103.04308}{\textcolor{magenta}{[arXiv:2103.04308 [quant-ph]]}}.
 
 \bibitem{HGDR}
 J. Hietarinta, B. Grammaticos, B. Dorizzi and A. Ramani, 
 \emph{``Coupling constant metamorphosis
and duality between integrable Hamiltonian systems",} 
\href{https://journals.aps.org/prl/abstract/10.1103/PhysRevLett.53.1707}{Phys. Rev. Lett. {\bf 53}  (1984) 1707}.

\bibitem{KMP}
E. G. Kalnins, W. Miller Jr. and S. Post,
  \emph{``Coupling constant metamorphosis and Nth-order symmetries 
  in classical and quantum mechanics",} 
  \href{https://iopscience.iop.org/article/10.1088/1751-8113/43/3/035202/meta}{J. Phys. A  {\bf 43}  (2010) 035202}
  \href{https://arxiv.org/abs/0908.4393}{\textcolor{magenta}{[arXiv:0908.4393 [math-ph]]}}.
 
\bibitem{MatSal}
V. B. Matveev and M. A. Salle, \textsl{Darboux Transformations and Solitons } (Springer, Berlin, 1991).

\bibitem{AraPlyu}
A. Arancibia and M. S. Plyushchay,
  \emph{``Chiral asymmetry in propagation of soliton defects in
crystalline backgrounds",} 
\href{https://doi.org/10.1103/PhysRevD.92.105009}{Phys. Rev. D {\bf 92}  (2016) 105009}
\href{https://arxiv.org/abs/1507.07060}{\textcolor{magenta}{[arXiv:1507.07060 [hep-th]]}}.

\bibitem{RogSch}
C. Rogers and W. K. Schief,
\textsl{B\"acklund and Darboux Transformations: 
Geometry and Modern Applications in Soliton Theory}
(Cambridge Univ. Press, 2002).

\bibitem{CBT1}
L.~Inzunza, M.~Plyushchay, and A.~Wipf,
\emph{``Conformal bridge between asymptotic freedom and confinement",}
\href{https://journals.aps.org/prd/abstract/10.1103/PhysRevD.101.105019}{Phys. Rev. D \textbf{101}  (2020) 105019}
\href{https://arxiv.org/abs/1912.11752}{\textcolor{magenta}{[arXiv:1912.11752 [hep-th]]}}. 

\bibitem{CBT2}
L.~Inzunza, M.~S.~Pyushchay and A.~Wipf, 
\emph{``Hidden symmetry and (super)conformal mechanics in a monopole background"},
\href{https://link.springer.com/article/10.1007/JHEP04(2020)028}{JHEP \textbf{04}  (2020) 028}
\href{https://arxiv.org/abs/2002.04341}{\textcolor{magenta}{[arXiv:2002.04341 [hep-th]]}}.

\bibitem{CBT3}
L.~Inzunza and M.~S.~Pyushchay, 
\emph{``Conformal bridge in a cosmic string background"},
\href{https://link.springer.com/article/10.1007/JHEP05(2021)165}{JHEP \textbf{05}   (2021) 165}
\href{https://arxiv.org/abs/2012.04613}{\textcolor{magenta}{[arXiv:2012.04613 [hep-th]]}}.


\bibitem{CBT4}
L.~Inzunza and M.~S.~Pyushchay, 
\emph{``Dynamics, symmetries, anomaly and vortices in a rotating cosmic string background"},
\href{https://link.springer.com/article/10.1007/JHEP01(2022)179}{JHEP   \textbf{01}  (2022) 179}
\href{https://arxiv.org/abs/2109.05161}{\textcolor{magenta}{[arXiv:2109.05161 [hep-th]]}}.

\bibitem{AchLiv1}
J. B. Achour and  E. R. Livine, 
\emph{``Symmetries and conformal bridge in Schwarschild-(A)dS black hole mechanics"},
\href{https://link.springer.com/article/10.1007/JHEP12(2021)152}{JHEP {\bf 12}  (2021) 152}
\href{https://arxiv.org/abs/2110.01455}{\textcolor{magenta}{[arXiv:2110.01455 [gr-qc]]}}.

\bibitem{AchLiv2}
J. B. Achour,  E. R. Livine, S. Mukohyama and J.-P. Uzan,
 \emph{``Hidden symmetry of the static response of black holes: applications to Love numbers"},
 \href{https://link.springer.com/article/10.1007/JHEP07(2022)112}{JHEP {\bf 07}  (2022) 112}
 \href{https://arxiv.org/abs/2202.12828}{\textcolor{magenta}{[arXiv:2202.12828 [gr-qc]]}}.

\bibitem{Bender}
C. M. Bender,  \textsl{PT symmetry in Quantum and Classical Physics} (World Scientific, 2019).

\bibitem{Dyson}
F. J. Dyson, \emph{``Thermodynamic behavior of an ideal ferromagnet"},
\href{https://journals.aps.org/pr/abstract/10.1103/PhysRev.102.1230}{Phys. Rev. {\bf 102}  (1956) 1230}.



\bibitem{CBT5}
L.~Inzunza and M.~S.~Plyushchay, 
\emph{``Conformal bridge transformation and PT symmetry"},
\href{https://iopscience.iop.org/article/10.1088/1742-6596/2038/1/012014}{J. Phys.: Conf. Ser. {\bf 2038}  (2021) 012014}
\href{https://arxiv.org/abs/2104.08351}{\textcolor{magenta}{[arXiv:2104.08351 [hep-th]]}}.


\bibitem{CBT6}
L.~Inzunza and M.~S.~Plyushchay, 
\emph{``Conformal bridge transformation, PT- and super- symmetry"},
\href{https://link.springer.com/article/10.1007/JHEP08(2022)228}{JHEP \textbf{08}   (2022) 228}
\href{https://arxiv.org/abs/2112.13455}{\textcolor{magenta}{[arXiv:2112.13455 [hep-th]]}}.

\bibitem{CBT7}
L.~Inzunza and M.~S.~Plyushchay, 
\emph{``Conformal generation of an exotic rotationally invariant harmonic oscillator"},
\href{https://journals.aps.org/prd/abstract/10.1103/PhysRevD.103.106004}{Phys. Rev. D \textbf{103}  (2021) 106004}
\href{https://arxiv.org/abs/2103.07752}{\textcolor{magenta}{[arXiv:2103.07752 [quant-ph]]}}.

\bibitem{ASW}
D. Arovas, J. R. Schrieffer, and F. Wilczek, 
\emph{``Fractional Statistics and the Quantum Hall Effect"}, 
\href{https://journals.aps.org/prl/abstract/10.1103/PhysRevLett.53.722}{Phys. Rev. Lett. {\bf 53}  (1984) 722}.

\bibitem{NSSFS}
C. Nayak, S. H. Simon, A. Stern, M. Freedman, and S. D. Sarma,
\emph{``Non-Abelian anyons and topological quantum computation"},
\href{https://journals.aps.org/rmp/abstract/10.1103/RevModPhys.80.1083}{Rev. Mod. Phys. {\bf  80}  (2008) 1083}
\href{https://arxiv.org/abs/0707.1889}{\textcolor{magenta}{[arXiv:0707.1889 [cond-mat.str-el]]}}.

\bibitem{LeiMyr}
J. M. Leinaas and J. Myrheim, \emph{``On the theory of identical particles"},
\href{https://link.springer.com/article/10.1007/BF02727953}{Nuovo Cim. B {\bf 37}  (1977) 1}.

\bibitem{Wilc1}
F. Wilczek, 
 \emph{``Quantum Mechanics of Fractional-Spin Particles"},
 \href{https://journals.aps.org/prl/abstract/10.1103/PhysRevLett.49.957}{Phys. Rev. Lett. {\bf 49}  (1982) 957}.

\bibitem{Wilc2}
F. Wilczek, \textsl{Fractional Statistics and Anyon Superconductivity} (World Scientific, Singapore, 1990).

\bibitem{Sefr}
S.  Serfaty,  \textsl{Coulomb Gases and
Ginzburg-Landau Vortices} (European Mathematical Society, 2015).

\bibitem{ChiHanMou}
R.~Y.~Chiao, A.~Hansen, and A.~A.~Moulthrop, 
\emph{``Fractional statistics of the vortex in two-dimensional superfluids"},
\href{https://journals.aps.org/prl/abstract/10.1103/PhysRevLett.54.1339}{Phys. Rev. Lett \textbf{54}  (1985) 1339}.

\bibitem{HalWu}
F.~D.~M.~Haldane and Yong-Shi Wu, 
\emph{``Quantum dynamics and statistics of vortices in two-dimensional superfluids"},
\href{https://journals.aps.org/prl/abstract/10.1103/PhysRevLett.55.2887}{Phys. Rev. Lett \textbf{55}    (1985) 2887}.


\bibitem{Penna}
V.~Penna, 
\emph{``Quantum dynamics of two-dimensional vortex pairs with arbitrary total vorticity"},
\href{https://journals.aps.org/prb/abstract/10.1103/PhysRevB.59.7127}{Phys. Rev. B \textbf{59} (1999)  7127}
\href{https://arxiv.org/abs/cond-mat/9909320}{\textcolor{magenta}{[arXiv:cond-mat/9909320]}}.

\bibitem{MalSta}
J. Maldacena and D. Stanford,
 \emph{``Remarks on the Sachdev-Ye-Kitaev model"},
\href{https://journals.aps.org/prd/abstract/10.1103/PhysRevD.94.106002}{Phys. Rev. D {\bf 94}  (2016) 106002}
 \href{https://arxiv.org/abs/1604.07818}{\textcolor{magenta}{[arXiv:1604.07818 [hep-th]]}}.
 
 \bibitem{MalStaYan}
 J.  Maldacena, D.  Stanford,  and Z. Yang, 
 \emph{``Conformal symmetry and its breaking in two-dimensional nearly anti-de Sitter space"},
 \href{https://academic.oup.com/ptep/article/2016/12/12C104/2624100?login=false}{Prog. 
 Theor. Exp. Phys. {\bf 2016}   (2016) 12C104}
 \href{https://arxiv.org/abs/1606.01857}{\textcolor{magenta}{[arXiv:1606.01857 [hep-th]]}}.

\bibitem{MerTurVer}
T. G. Mertens, G. J. Turiaci, and H. L.  Verlinde, 
  \emph{``Solving the Schwarzian via the conformal bootstrapm"}, 
 \href{https://link.springer.com/article/10.1007/JHEP08(2017)136}{ J. High Energ. Phys. {\bf 08} (2017) 136}
 \href{https://arxiv.org/abs/1705.08408}{\textcolor{magenta}{[arXiv:1705.08408 [hep-th]]}}.

\bibitem{GanPol}
S. Ganeshan and A. P. Polychronakos,
\emph{``Lyapunov exponents and entanglement entropy transition
on the noncommutative hyperbolic plane"}, 
\href{https://scipost.org/SciPostPhysCore.3.1.003}{SciPost Phys. Core {\bf 3} (2020)  003}
\href{https://arxiv.org/abs/1912.10805}{\textcolor{magenta}{[arXiv:1912.10805 [hep-th]]}}.



\bibitem{LeinMyrh}
J.~M.~Leinaas and and J.~Myrheim, 
\emph{``Intermediate statistics for vortices in superfluid films"},
\href{https://journals.aps.org/prb/abstract/10.1103/PhysRevB.37.9286}{Phys. Rev. B \textbf{37} (1988) 9286}.

\bibitem{Leinaas}
J.~M.~Leinaas, 
\emph{``Quantized vortex motion and the motion of charged particles in a strong magnetic field"},
\href{https://www.sciencedirect.com/science/article/abs/pii/000349169090326J}{Annals Phys. \textbf{198}  (1990) 24}.

\bibitem{non-Lan1}
C.~Duval and P.~A.~Horv\'athy,
\emph{``The exotic Galilei group and the "Peierls substitution""},
\href{https://www.sciencedirect.com/science/article/abs/pii/S0370269300003415?via%3Dihub}{Phys. Lett. B \textbf{479}  (2000) 284}
\href{https://arxiv.org/abs/hep-th/0002233}{\textcolor{magenta}{[arXiv:hep-th/0002233]}}.

\bibitem{non-Lan2}
J.~Gamboa, M.~Loewe and J.~C.~Rojas, 
\emph{``Noncommutative quantum mechanics"},
\href{https://journals.aps.org/prd/abstract/10.1103/PhysRevD.64.067901}{Phys. Rev. D \textbf{64}  (2001) 067901}
\href{https://arxiv.org/abs/hep-th/0010220}{\textcolor{magenta}{[arXiv:hep-th/0010220]}}.

\bibitem{non-Lan2+}
V. P. Nair and  A. P. Polychronakos,
\emph{``Quantum mechanics on the noncommutative plane and sphere"},
\href{https://www.sciencedirect.com/science/article/abs/pii/S0370269301003392}{Phys. Lett. B {\bf 505} (2001) 267}
\href{https://arxiv.org/abs/hep-th/0011172}{\textcolor{magenta}{[arXiv:hep-th/0011172]}}.


\bibitem{HorDuv}
C.~Duval and P.~A.~Horv\'athy,
\emph{``Exotic Galilean symmetry in the noncommutative plane, and the Hall effect"}, 
\href{https://iopscience.iop.org/article/10.1088/0305-4470/34/47/314}{J. Phys. A {\bf 34}
 (2001)  10097}
 \href{https://arxiv.org/abs/hep-th/0106089}{\textcolor{magenta}{[arXiv:hep-th/0106089]}}.



\bibitem{KNP}
D. Karabali, V. P. Nair and A. P. Polychronakos,
\emph{``Spectrum of Schroedinger 
field in a noncommutative magnetic monopole"},
\href{https://www.sciencedirect.com/science/article/abs/pii/S0550321302000627?via%3Dihub}{	Nucl. Phys. B {\bf  627}   (2002) 565}
 \href{https://arxiv.org/abs/hep-th/0111249}{\textcolor{magenta}{[arXiv:hep-th/0111249]}}.    
 

\bibitem{non-Lan3}
P.~A.~Horvathy, 
\emph{``The non-commutative Landau problem"},
\href{https://www.sciencedirect.com/science/article/abs/pii/S0003491602962718}{Annals Phys. \textbf{299}  (2002) 128}
\href{https://arxiv.org/abs/hep-th/0201007}{\textcolor{magenta}{[arXiv:hep-th/0201007]}}.


\bibitem{non-Lan4}
P.~A.~Horvathy and M.~S.~Plyushchay, 
\emph{``Nonrelativistic anyons, exotic Galilean symmetry and
noncommutative plane"},
\href{https://iopscience.iop.org/article/10.1088/1126-6708/2002/06/033}{JHEP \textbf{06} (2002) 033}
\href{https://arxiv.org/abs/hep-th/0201228}{\textcolor{magenta}{[arXiv:hep-th/0201228]}}.

\bibitem{Ncc}
P.~A.~Horvathy, 
\emph{``Non-commuting coordinates in vortex dynamics and in the Hall effect, related to "exotic" Galilean symmetry"},
\href{https://www.worldscientific.com/doi/abs/10.1142/9789812704467_0026}{In: Nonlinear Physics: Theory and Experiment. II, pp. 186--192},
Eds. M. J. Ablowitz et al. (World Sci. Pub., Sinhapore, 2003)
\href{https://arxiv.org/abs/hep-th/0207075}{\textcolor{magenta}{[arXiv:hep-th/0207075]}}.

\bibitem{non-Lan5}
P.~A.~Horvathy and M.~S.~Plyushchay, 
\emph{``Nonrelativistic anyons in external electromagnetic field"},
\href{https://www.sciencedirect.com/science/article/abs/pii/S0550321305001719}{Nucl. Phys. B \textbf{714}  (2005) 269}
\href{https://arxiv.org/abs/hep-th/0502040}{\textcolor{magenta}{[arXiv:hep-th/0502040]}}.


\bibitem{VPNair}
V. P.  Nair, 
\emph{``Noncommutative mechanics, Landau levels, twistors and Yang-Mills amplitudes"}, 
\href{https://link.springer.com/chapter/10.1007/3-540-33314-2_3}{ 
Lect. Notes Phys. {\bf 698}  (2006) 97}
\href{https://arxiv.org/abs/hep-th/0506120}{\textcolor{magenta}{[arXiv:hep-th/0506120]}}.




\bibitem{non-Lan6}
M.~del Olmo, M.~S.~Plyushchay, 
\emph{``Electric Chern-Simons term, enlarged exotic Galilei symmetry and noncommutative plane"},
\href{https://www.sciencedirect.com/science/article/abs/pii/S0003491606000650}{Annals Phys. \textbf{321}  (2006) 2830}
\href{https://arxiv.org/abs/hep-th/0508020}{\textcolor{magenta}{[arXiv:hep-th/0508020]}}.

\bibitem{non-Lan6+}
P. D. Alvarez, J. Gomis, K. Kamimura, and M.  S. Plyushchay,
\emph{``Anisotropic harmonic oscillator, non-commutative Landau problem 
and exotic Newton-Hooke symmetry"},
\href{https://www.sciencedirect.com/science/article/pii/S0370269307015353}{Phys. Lett. B {\bf 659}  (2008) 906}
\href{https://arxiv.org/abs/0711.2644}{\textcolor{magenta}{[arXiv:0711.2644 [hep-th]]}}.

\bibitem{non-Lan7}
P. D. Alvarez, J. L. Cort\'es, P. A. Horv\'athy and M. S. Plyushchay,
\emph{``Super-extended noncommutative Landau problem and conformal symmetry"},
\href{https://iopscience.iop.org/article/10.1088/1126-6708/2009/03/034}{JHEP {\bf 03}  (2009) 034}
\href{https://arxiv.org/abs/0901.1021}{\textcolor{magenta}{[arXiv:0901.1021 [hep-th]]}}.

\bibitem{ArnoldKhesin}
V. I. Arnold and B. A. Khesin,  \textsl{Topological Methods in Hydrodynamics},
Applied Mathematical Sciences, vol. {\bf 125} (Springer-Verlag, New York, 1998).

\bibitem{Newton}
P. K. Newton,
 \textsl{The N-vortex Problem: Analytical Techniques}
(Springer-Verlag, New York, 2001).



 \bibitem{Kirchh}
G. Kirchhoff,   \textsl{Vorlesungen \"uber mathematische Physik, Vol. 1. Mechanik}. 
 Teubner, Leipzig (1876). 
 
\bibitem{DJT}
G. V. Dunne, R. Jackiw and  C.A. Trugenberger,
 \emph{````Topological" (Chern-Simons) quantum mechanics"}, 
 \href{https://journals.aps.org/prd/abstract/10.1103/PhysRevD.41.661}{Phys. Rev. D {\bf 41}   (1990) 661}.


 \bibitem{Galazh}
 A. Galajinsky, 
 \emph{``Generalised point vortices on a plane"},
 \href{https://www.sciencedirect.com/science/article/pii/S0370269322002532?via%3Dihub}{Phys.  Lett. B
{\bf 829}   (2022) 137119}
\href{https://arxiv.org/abs/2203.00273}{\textcolor{magenta}{[arXiv:2203.00273 [hep-th]]}}.
 
 
 \bibitem{MarcClaudio}
 M. Henneaux and C. Teitelboim,
 \textsl{Quantization of Gauge Systems} (Princeton Univ. Press, 1992).
 
 \bibitem{JackNair}
 R. Jackiw, V. P. Nair,  \emph{``Anyon spin and the exotic central extension of the planar
Galilei group"}, 
 \href{https://www.sciencedirect.com/science/article/abs/pii/S0370269300003798}{Phys. Lett. B {\bf 480} (2000) 237}
 \href{https://arxiv.org/abs/hep-th/0003130}{\textcolor{magenta}{[arXiv:hep-th/0003130]}}.

\bibitem{AnyNCP}
P.~A.~Horvathy and M.~S.~Plyushchay, 
\emph{``Anyon wave equation and the noncommutative plane"},
\href{https://www.sciencedirect.com/science/article/pii/S0370269304008378}{Phys. Lett. B \textbf{595}  (2004) 547}
\href{https://arxiv.org/abs/hep-th/0404137}{\textcolor{magenta}{[arXiv:hep-th/0404137]}}.

\bibitem{CJP}
F. Correa, V. Jakubsky and M. S. Plyushchay, \emph{``PT-symmetric invisible defects and
confluent Darboux-Crum transformations"},
 \href{https://journals.aps.org/pra/abstract/10.1103/PhysRevA.92.023839}{ Phys. Rev. A {\bf 92}  (2015) 023839}
\href{https://arxiv.org/abs/1506.00991}{\textcolor{magenta}{[arXiv:1506.00991 [quant-ph]}}.

\bibitem{CarPlyu}
J. F. Cari\~nena and M. S. Plyushchay,
\emph{``Ground-state isolation and discrete flows in a
rationally extended quantum harmonic oscillator"},
\href{https://journals.aps.org/prd/abstract/10.1103/PhysRevD.94.105022}{Phys. Rev. D {\bf 94}  (2016) 105022}
\href{https://arxiv.org/abs/1611.08051}{\textcolor{magenta}{[arXiv:1611.08051 [hep-th]}}.

\bibitem{Dirac}
P. A. M. Dirac, 
\emph{``Forms of relativistic dynamics"}, 
\href{https://journals.aps.org/rmp/abstract/10.1103/RevModPhys.21.392}{Rev. Mod. Phys. {\bf 21}  (1949) 392}.




\bibitem{CFTbook}
Ph. Francesco , P.  Mathieu  and D.  Senechal, 
 \textsl{Conformal Field Theory}, (Springer, 1997).


 

\bibitem{Fractons}
D. Doshi and A.  Gromov,
 \emph{``Vortices as fractons"},  
\href{https://www.nature.com/articles/s42005-021-00540-4}{Commun.  Phys.  {\bf 4}  (2021) 44}.


\bibitem{ComHou}
A. Comtet and P. J. Houston, 
 \emph{``Effective action on the hyperbolic plane in a constant external
field"}, 
\href{https://pubs.aip.org/aip/jmp/article-abstract/26/1/185/226570/Effective-action-on-the-hyperbolic-plane-in-a?redirectedFrom=fulltext}{J. 
Math. Phys. {\bf 26}  (1985) 185}. 


\bibitem{Comtet}
A. Comtet,
\emph{``On the Landau levels on the hyperbolic plane"},
\href{https://www.sciencedirect.com/science/article/abs/pii/0003491687900984}{Ann. Physics {\bf 173}  (1987) 185}.


 \bibitem{MPAnn}
 M.  S. Plyushchay, 
 \emph{``Deformed Heisenberg algebra, fractional spin fields, and supersymmetry without fermions"}, 
 \href{https://www.sciencedirect.com/science/article/abs/pii/S0003491696900123}{Annals Phys. {\bf 245}  (1996) 339}
 \href{https://arxiv.org/abs/hep-th/9601116}{\textcolor{magenta}{[arXiv:hep-th/96011161]}}.
 
 
 

\end{thebibliography}
 \end{document}